# Universal imprinting of chirality with chiral light by employing plasmonic metastructures

Oscar Ávalos-Ovando[1,2*], Veronica A. Bahamondes Lorca[3,4], Lucas V. Besteiro[5], Artur Movsesyan[1,6], Zhiming Wang[6*], Gil Markovich[7*], and Alexander O. Govorov[1,2*]

[1] Department of Physics and Astronomy, Ohio University, Athens, Ohio 45701, United States

[2] Nanoscale and Quantum Phenomena Institute, Ohio University, Athens, Ohio 45701, United States

[3] Edison Biotechnology Institute, Ohio University, Athens 45701, OH

[4] Departamento de Tecnología médica, Facultad de Medicina, Universidad de Chile, Santiago, Chile

[5] CINBIO, Universidade de Vigo, 36310 Vigo, Spain

[6] Institute of Fundamental and Frontier Sciences, University of Electronic Science and Technology of China, Chengdu 610054, China

[7] School of Chemistry, Raymond and Beverly Sackler Faculty of Exact Sciences, Tel Aviv University, Tel Aviv, 6997801 Israel

* Corresponding authors: oa322219@ohio.edu, zhmwang@uestc.edu.cn, gilmar@tauex.tau.ac.il, govorov@ohio.edu

# Abstract

Chirality, either of light or matter, has proved to be very practical in biosensing and nanophotonics. However, the fundamental understanding of its temporal dynamics still needs to be discovered. A realistic setup for this are the so-called metastructures, since they are optically active and are built massively, hence rendering an immediate potential candidate. Here we propose and study the electromagnetic-optical mechanism leading to chiral optical imprinting on metastructures. Induced photothermal responses create anisotropic permittivity modulations, different for left or right circularly polarized light, leading to temporal-dependent chiral imprinting of hot-spots, namely 'imprinting of chirality.' The above effect has not been observed yet, but it is within reach of modern experimental approaches. The proposed nonlinear chiroptical effect is general and should appear in any anisotropic material; however, we need to design a particular geometry for this effect to be strong. These new chiral time-dependent metastructures may lead to a plethora of applications.

## I. Introduction

Plasmonic nano-systems have been largely studied over the last years, since these systems exhibit unusually strong near-field and electromagnetic responses that enable efficient material-light manipulation, leading to promising ultrafast nanophotonic applications.[1] Additionally, these nano-systems generate heat efficiently in the presence of light and are even stronger when this is at the plasmon resonance frequency. Furthermore, a new tunable-knob in plasmonics has emerged over the past decade, chiral plasmonics, which can be structural or electromagnetic.[2–4] Specifically, one branch of plasmonics is focused on the so-called metastructures (or metamaterials), which are repeated patterns of some material/shape supported by a substrate that collectively create a new resonant structure.[5] Engineering these metastructures could lead to new ultrafast light manipulation devices and applications.[6,7]

When this concept of substrate-supported superlattices is combined with chirality, it allows to reduce the effective dimensionality of 3D objects, creating a quasi-2D lattice, in order to obtain larger chiral dissymmetry factors.[8] Recently, it was shown that when such arrangement of chiral nanoparticles is illuminated with circular polarized light (CPL), collective plasmon oscillations create a strong circular dichroism (CD)[9], which can be further enhanced if there are chiral molecules in the media, leading to an ultrasensitive chirality detector[10,11]. Also, achiral gold nanoparticles arranged in a tetrahedral fashion over a substrate, will become a chiral *pinwheel* superlattice,[12] and most recently, collective resonant metasurfaces will support chiral bound states in the continuum which enhances CD signals,[13,14] among other prominent examples[8,15–20].

Whereas chiral metastructures have been previously studied in the continuous wave (CW) regime,[21,22] especially when made of noble metals such as gold and silver, there are few to none reports on how chiral photothermal responses behave at ultrafast time scales. Here we study both chiral and achiral metastructures, and how they respond to ultrashort CPL illumination pulses, by calculating their thermo-temporal responses, both spatially and spectrally. We use both the one- (1T) and the two-temperature (2T) model for the thermal dynamics, but we modify them in order to account anisotropic changes[23–26] of the material's permittivity after illumination, and discuss the differences between both models. We find that CPL pulses induced new near-field chiral patterns in different plasmonics hot-spots of each structure, at different time scales, which we identify and classify. We find that this effect is rather universal at every structure we studied, which is why we termed the effect 'chirality imprinting'. We also discuss why the mostly used thermal 1T model[27] cannot fully describe the photothermal effects associated to chirality imprinting, such that the 2T model becomes necessary, as recently shown for some achiral metastructures.[28,29] Our findings could be tested with time-resolved photothermal CD spectroscopy, since prominent asymmetries should be notable in their local temperature maps.

## II. Formalism

For our computational models and numerical simulations, we consider the interaction of the gold structures with their environment media, the latter set as either water or glass as it will be specified in/for every case. We irradiate the gold nanostructures with CPL in two regimes, either the static CW (linear response) or the temporal pulsed excitation, and then for the latter we study the photothermal induced heat propagation due to photoinjected carriers[30] with two models: a 1T model which solves the lattice temperature $T_l$ in the system, and a 2T model which solves both the electronic temperature $T_e$ and the lattice temperature $T_l$ simultaneously. Changes in temperature will affect the material's permittivity at different times, in the so-called symmetry-breaking window, creating anisotropic permittivity patterns, before reaching thermal relaxation.[23–26]

### 1. Formalism for the photothermal response.

Our classical electromagnetic simulations for the optically-excited Au nanoparticles and metastructures are performed via the Finite Elements Method with the COMSOL Multiphysics simulations software. We illuminate the metastructures with linear and/or CPL and calculate the photothermal response of the system. The incident electromagnetic field is defined as: $\vec{E}_{ext} = \text{Re}\left[\vec{E}_0 e^{i\omega t}\right]$. The calculations provide principally $Q_h$, the total power dissipation density (in units of W/m³), from which far-field quantities such the absorption ($\sigma_{abs}$), scattering ($\sigma_{scat}$), and extinction ($\sigma_{ext}$) optical cross-sections (related by $\sigma_{ext} = \sigma_{abs} + \sigma_{scat}$) can be calculated by solving the Maxwell's equations within a classical framework. In particular, the formalism for the absorption cross-section is based on the following equations:

$$Q_{abs} = -\text{Im}(\varepsilon_{metal})\frac{\varepsilon_0 \omega}{2}\int dV\, \vec{E}_\omega \cdot \vec{E}_\omega^*, \qquad (1)$$

$$\sigma_{abs} = \frac{Q_{abs}}{I_0}, \qquad (2)$$

where $Q_{abs}$ is the absorbed power by the system, $\varepsilon_{metal}$ is the dielectric constant of the metal nanocrystal (NC), $\omega$ is the angular frequency of the incident light, $\vec{E}_\omega$ is the complex electric field inside the metal, and $I_0$ is the photon flux magnitude (in the main text referred only as intensity for simplicity), given by

$$I_0 = \frac{c_0 \varepsilon_0 \varepsilon_{med}^{1/2}}{2}\left|\vec{E}_0\right|^2, \qquad (3)$$

where $\varepsilon_{med}$ is the dielectric constant of the medium, $c_0$ is the speed of light in vacuum, $\varepsilon_0$ is the vacuum permittivity, and $\left|\vec{E}_0\right|$ is the electric field magnitude of the incident electromagnetic wave. For $\varepsilon_{metal}$ we use a gold Drude-like dielectric constant as defined below in Section II.2.

In general in this work, we illuminate the metastructures from above ($k \parallel -z$) with an incident CPL light polarization ($\alpha =$ LCP, RCP), and the photoinduced spatial temperature changes with respect to the background ($T_0 \equiv 293.15$ K) are calculated via two models, as explained later in Section II.3.

We calculate optical absorptions of each structure, defined as

$$A_\alpha = \frac{1}{P}\int Q_h \, dV, \tag{4}$$

where $Q_h$ is the total power dissipation density (in units of W/m³), $P$ is the irradiated power density ($P = I_0 \cdot p_x \cdot p_y$, and $p_{x,y}$ the periodic dimension of each unit cell), and lastly the integral is taken either over the entire Au antenna's, the Au mirror's, or the whole system's volume. As each metastructure presents a thick gold layer below, it is expected to behave as a so-called metamaterial perfect absorber, showing nearly no transmission (so $A + R = 1$). Then, the chirality of the system is calculated via the circular dichroism (CD) and the dissymmetry factor (g-factor), defined as

$$CD_A = A_{LCP} - A_{RCP}, \tag{5}$$

and

$$g_A = \frac{A_{LCP} - A_{RCP}}{(A_{LCP} + A_{RCP})/2}. \tag{6}$$

## 2. Temperature dependent Drude model

Drude-like systems were simulated with custom-made dielectric constants $\varepsilon_{\text{metal}}$, which can be tailored to highlight the relevant physics similar as gold, as usually done in the literature when one wants to avoid gold interband effects. The relative permittivity within the Drude model is given by

$$\varepsilon_{\text{metal,bulk}}(\omega) = \varepsilon_{b,\text{Drude}} - \frac{\omega_{p0}^2}{\omega \cdot (\omega - i\gamma_{D0})}, \tag{7}$$

where $\varepsilon_{b,\text{Drude}}$ is the long-wavelength background dielectric constant, $\omega_{p0}$ is the plasma frequency, and $\gamma_{D0}$ is the damping coefficient. In this expression, $\omega$ is given in units of eV, as $\omega = 2\pi\hbar c/\lambda$, and the zero subindex stands for non-temperature dependent. Table S1 summarizes the Drude parameters, taken from Ref. [31]. In Figure S1 one can see the successful long-wavelength comparison between this Drude model and realistic gold dielectric functions found in the literature[32,33], the latter widely used for simulating real materials.

When temperature is considered, it is known that the lattice temperature $T_l$ will change the metal's permittivity due to free electrons (Drude-Sommerfeld permittivity) via two mechanisms: the plasma frequency and the Drude damping[34–36]. Then, the T-dependent Drude permittivity can be written as

$$\varepsilon_{\text{metal, bulk}}(\omega, T) = \varepsilon_{b,\text{Drude}}(T) - \frac{\omega_p^2(T)}{\omega(\omega - i\gamma_D(T))}, \tag{8}$$

where the plasma frequency is modified as an isotropic and adiabatic process as

$$\omega_p(T) = \sqrt{\frac{4\pi e^2 n_0(T)}{m}} \approx \omega_{p0}(T_0) - \frac{\alpha_V}{2}\omega_{p0}(T_l - T_0),$$

with $e$ is the electron charge, $n_0(T)$ is the T-dependent number of electrons, and $m$ is the electron's mass. Also, $\alpha_V \equiv 3\alpha_L$ is the coefficient of volume thermal expansion and $T_0 \equiv 293.15$ K is the room temperature[37]. The damping coefficient is modified as

$$\gamma_D(T) = \gamma_{D0} + \beta(T_l - T_0)\gamma_{e-ph}$$

where $\beta$ is a Holstein's model estimated constant, and $\gamma_{e-ph}$ is the electron-phonon scattering driven T-dependent contribution to the Drude damping. Also, even though $\varepsilon_{b,Drude}(T)$ in principle is T-dependent we take it as constant without loss of generality. Finally, these three T-dependent parameters are shown in Fig. S1b, where a direct comparison to recent experimental values[38] has been done finding excellent agreement with our estimations. The corresponding dielectric constants given by Eq. 8 are shown in Fig. S1c.

### 3. Thermal heat-diffusion models

Here we provide details of the two heating models used in our work, the one-temperature model (1T model) which solves the lattice temperature ($T_l$) after a pulsed illumination, and the two-temperature model (2T model) which solves the lattice ($T_l$) and electron ($T_e$) temperatures after a pulsed illumination, when a metal nanostructure (gold) is embedded in some media (we use both water and glass). The electronic temperature $T_e$ is set to exist only in the metal, i.e. the gold, so it is fully isolated from the media. On the other hand, the lattice temperature $T_l$ is set to exist in both the gold and the media. As such, there is electronic-lattice (electron-phonon) heat exchange within gold, which is later relaxed as lattice-lattice (phonon-phonon) heat exchange between gold and media. At the same time, the outermost boundaries are always kept fixed at $T_0 \equiv 293.15$ K (see Fig. S2).

Both models take as an input the 3D spatial absorbed power $q(\vec{r},t)$, defines as

$$q(\vec{r},t) = \langle \vec{j}(\vec{r},t) \cdot \vec{E}(\vec{r},t) \rangle_t = Q(\vec{r}) \cdot F_{pulse}(t) \qquad (9)$$

where $F_{pulse}(t)$ is a Gaussian pulse defined as

$$F_{pulse}(t) = \exp\left(-\frac{(t-t_0)^2}{2\sigma^2}\right) \qquad (10)$$

with $t_0 = 1$ ps is the time of the pulse emission, and $\sigma = FWHM / 2.3548$, with $FWHM = 100$ fs the duration of the pulse. In general, $F_{pulse}(t)$ is a dimensionless function describing any pulse profile,[39] which in the CW regime or in cases of very long pulses $F_{pulse}(t) \equiv 1$, which is the so-called steady-state regime. The quantity $Q_h(\vec{r})$ is the steady-state 3D map of the light dissipation throughout and the subsequent response recorded as a spatial absorbed heat in all the domains, both metal and media, which is obtained directly from Comsol, via a variable called total power dissipation density, which is now used as input for the single (1T model) o coupled (2T model) thermal modules.

### 3.1. One-Temperature model (1T model).

In this model we only solve $T_l$ throughout all the domains, but the one creating the heating is only the metal, which in our case is gold. Under the pulsed excitation previously described, the photo-temperature distribution is computed from the local thermal heat-diffusion equation.[40]

$$\rho(\vec{r})\,c(\vec{r})\frac{\partial T_l(\vec{r},t)}{\partial t} = \vec{\nabla}\cdot\left[k(\vec{r})\left(\vec{\nabla}T_l(\vec{r},t)\right)\right] + Q_h(\vec{r},t) \qquad (11)$$

where $T_l(\vec{r},t)$ is the local temperature distribution as a function of the coordinate $\vec{r}$ and the time $t$, and $Q_h(\vec{r},t)$ is the pulsed-induced heating term previously described as $Q_h(\vec{r})$ in the static regime. The function $T_l(\vec{r},t)$ is the excess temperature induced by light in the vicinity of gold. $\rho(\vec{r})$, $c(\vec{r})$ and $\kappa(\vec{r})$ are the mass density, the specific heat, and the thermal conductivity of each material, respectively. For simplicity, we assume that the media around the gold is a uniform media.

### 3.2. Two-temperature model (2T model).

We follow the dual-parabolic two-step model proposed in Refs. [41,42] to include the heat transfer contributions within electrons and phonons, given by the coupled differential equations

$$C_e(T_e)\frac{\partial T_e}{\partial t} = \nabla\left[\kappa_e(T_e)\nabla T_e\right] - G[T_e - T_l] + Q_h(\vec{r},t)$$
$$C_l(T_l)\frac{\partial T_l}{\partial t} = \nabla\left[\kappa_l(T_l)\nabla T_l\right] + G[T_e - T_l] \qquad (12)$$

which couple electrons ($e$) and lattice ($l$) thermal contributions in the metal nanoparticle. Both temperatures ($T_e$ and $T_l$) are function of both space and time so $T_e(\vec{r},t)$ and $T_l(\vec{r},t)$, and are coupled together via the time ($t$). They are also modulated by temperature-dependent quantities: the volumetric heat capacity $C(T)$, the thermal conductivity $\kappa(T)$, and the pulsed-induced heating term $Q_h(\vec{r},t)$ previously described. The $G$ term is the electron-phonon coupling constant, and in this work is taken as $2.78\times10^{16}$ W m$^{-3}$ K$^{-1}$, in good agreement with the literature in which varies for different structures between 1 and 4 in units of $\left[10^{16}\text{ W m}^{-3}\text{ K}^{-1}\right]$[43–46]. In this model the $T_e(\vec{r},t)$ is defined only in the gold, whereas $T_l(\vec{r},t)$ is defined everywhere (as in the 1T model), in this way the electronic energy is transferred to the vibrations of the lattice (i.e. the phonons), until thermal equilibrium is achieved.

In metals, the total thermal conductivity can be written as $\kappa(T) \approx a\kappa_e(T_e) + b\kappa_l(T_l)$, where $a$ and $b$ are constants. For gold, where most of the thermal conductivity comes from electrons

$a = 0.99$ and $b = 0.01^{41}$. For gold, the electronic heat capacity is defined with the Debye approximation, as

$$C_e(T_e) = \frac{\gamma \rho T_e}{M_W} \approx 71.4 T_e \left[\text{J m}^{-3}\text{ K}^{-1}\right],$$

where $\gamma$ is the Sommerfeld constant, $\rho$ is the gold density, and $M_W$ is the molar weight. $C_l(T_l)$ and $\kappa_l$ are taken from Comsol libraries.

### 4. CLP pump-probe model.

After either the 1T or the 2T models are applied, we record 3D maps of the system thermal response at every time, in particular $T_l(\vec{r},t)$ within the model we are using, such that the T-dependent Drude permittivity Eq 5 gains a $t$ dependence, so the permittivity reads

$$\varepsilon_{\text{metal, bulk}}(\omega, T_l^{CPL}) = \varepsilon_{\text{b,Drude}}(T_l^{CPL}(\vec{r},t)) - \frac{\omega_p^2(T_l^{CPL}(\vec{r},t))}{\omega(\omega - i\gamma_D(T_l^{CPL}(\vec{r},t)))} \tag{13}$$

where the superscript CPL was the incident light polarization (LCP, RCP). Depending which time $t$ and CPL polarization is chosen, $\varepsilon_{\text{metal, bulk}}(\omega, T_l^{CPL})$ will be the input permittivity which simulates a given CPL pumping, which we again recalculate the system CW response with either CPL probe. We then can define spatial distributions of the system chirality as

$$CD_{|E/E_0|} = \left(\frac{E}{E_0}\right)_{LCP} - \left(\frac{E}{E_0}\right)_{RCP} \tag{14}$$

### III. Chiral and Achiral Metamodels and Optics

Figure 1 shows the metastructures we study here, additionally to simple spherical gold nanoparticles (not shown) to help us gain understanding of the performance efficiency of our thermal models. The typical pump-probe experimental irradiation conditions here simulated consider CPL either $\alpha =$ LCP or RCP for either the pump and/or the probe, as shown in Fig. 1a. The metastructures are shown in Figs. 1b and 1c (see also Figures S2d and S2e for our periodic boundary conditions).

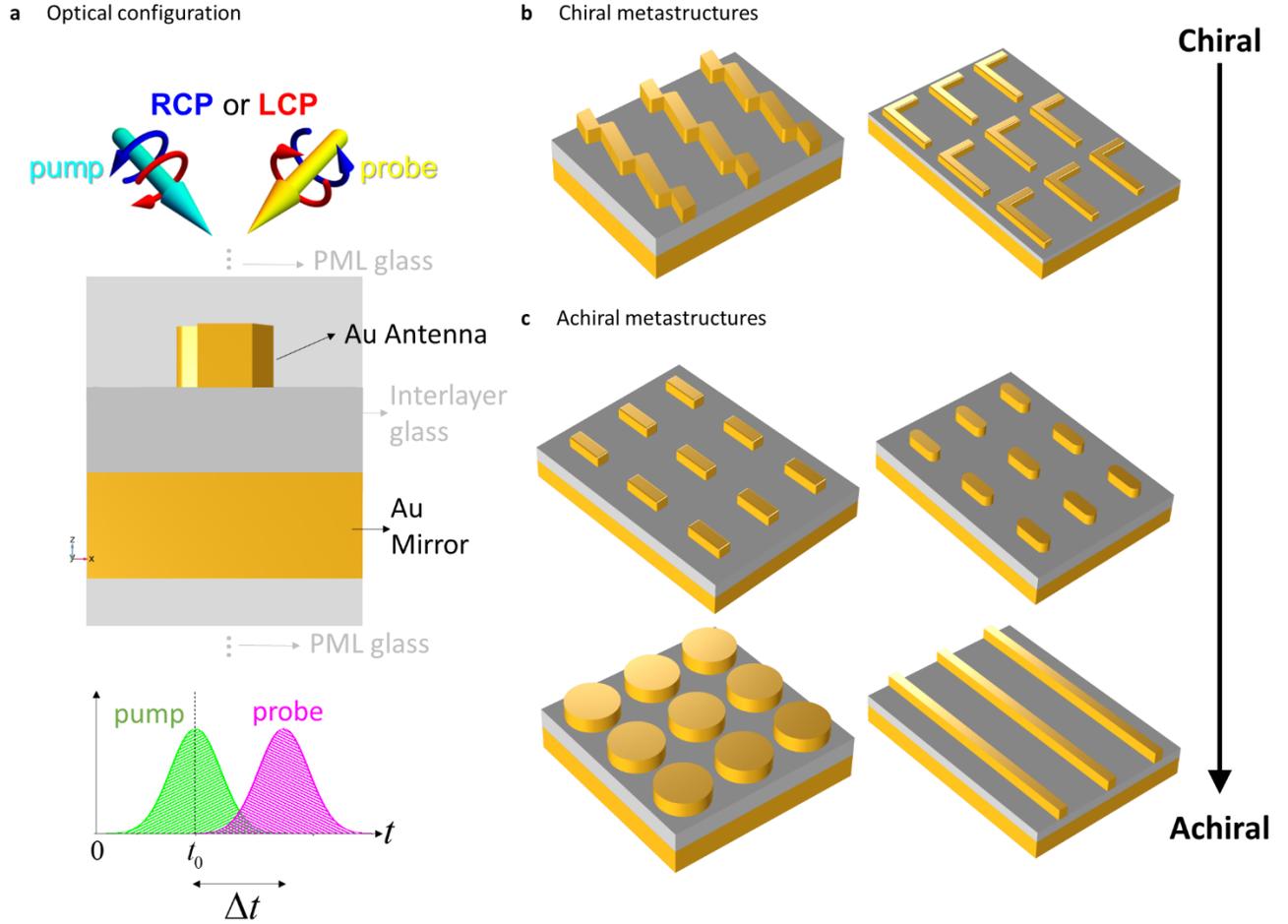

**Figure 1: Schematics of the metastructures.** (a) Unit cell side-view of a typical metastructure setup studied here: the golden regions are gold, and the gray regions are glass. The metastructure is illuminated from above with a pump CPL, which photoexcites it and induces a subsequent photoheating, and then probed by CPL at different time scales. Each structure considers infinite perfectly matched layer (PML) of glass towards both the +z and -z directions (gray dots), schematized only here but not shown in what follows for clarity (see sup. material, Fig. S13 for clarification). The bottom scheme at panel (a) shows the typical pump-probe temporal simulation, in where a pump beam excites the structures and after a time $\Delta t = t - t_0$ the probe beam measures the system's response. (b) the chiral metastructures L-shaped (left) and chiral wire (right), and (c) achiral metastructures disk and rod in the top row, and rectangular prism and achiral wire in the bottom row. Note that each geometry shown in b and c is a composition of 9 unit cells for better visualization of the metastructure, such that then every periodic structure is composed by infinite unit cells which are repeated using periodic boundary conditions throughout the XY plane. For details about each geometry see the supplemental material, figures S5 and S6.

We first validate our T-models with a simple Au nanosphere, as such experimental results are known. See Supplemental material and figures S3, S4 and S5 for full description. We find that typical electron-phonon relaxation times, $\tau_{e-ph}$, of ~1 ps, whereas our phonon-phonon relaxation times are in the order of several tens of ps, in well agreement with literature,[47–50] validating our models. Note that $\tau_{e-ph}$ largely depends on the irradiation intensity $I_0$, so a correct scale down needs to be considered when modelling with high intensities (see Figs. S5, S17 and S26 for estimations of such scaling). We then move to more complex metastructures, in order to account for typical experimental scenarios. In particular, we consider two realistic chiral metastructures

(an L-shaped[39,51] and a chiral wire[52,53]) and four realistic achiral metastructures (disk,[27,54,55] rod,[56–58] rectangular prism,[28,57] and achiral wire[24]), as shown in Fig. 1. In modeling the latter kind, we adopt a four-fold symmetric finite element approach,[59] in order to avoid any numerical artificial chirality (see Fig. S14). Then, the chirality of the system is calculated via the CD and the dissymmetry factor (g-factor), Eqs. 5 and 6, respectively. The different dimensions of each metastructure used in our simulations are specified and schematized in Figs. S5 and S6, where the purpose of each choice of sizes is twofold: for the sizes to be within experimental size and for the main plasmon to appear in the visible near-infrared range of the absorption spectra. The sizes of the Au mirror and the interlayer glass gap (between antenna and mirror) are chosen to be within typical experimental lithographic techniques, and their ratio to be good enough to yield a zero transmission through the system, so that A+R=1, with R being the optical reflectivity and A the absorption of Eq. 4.

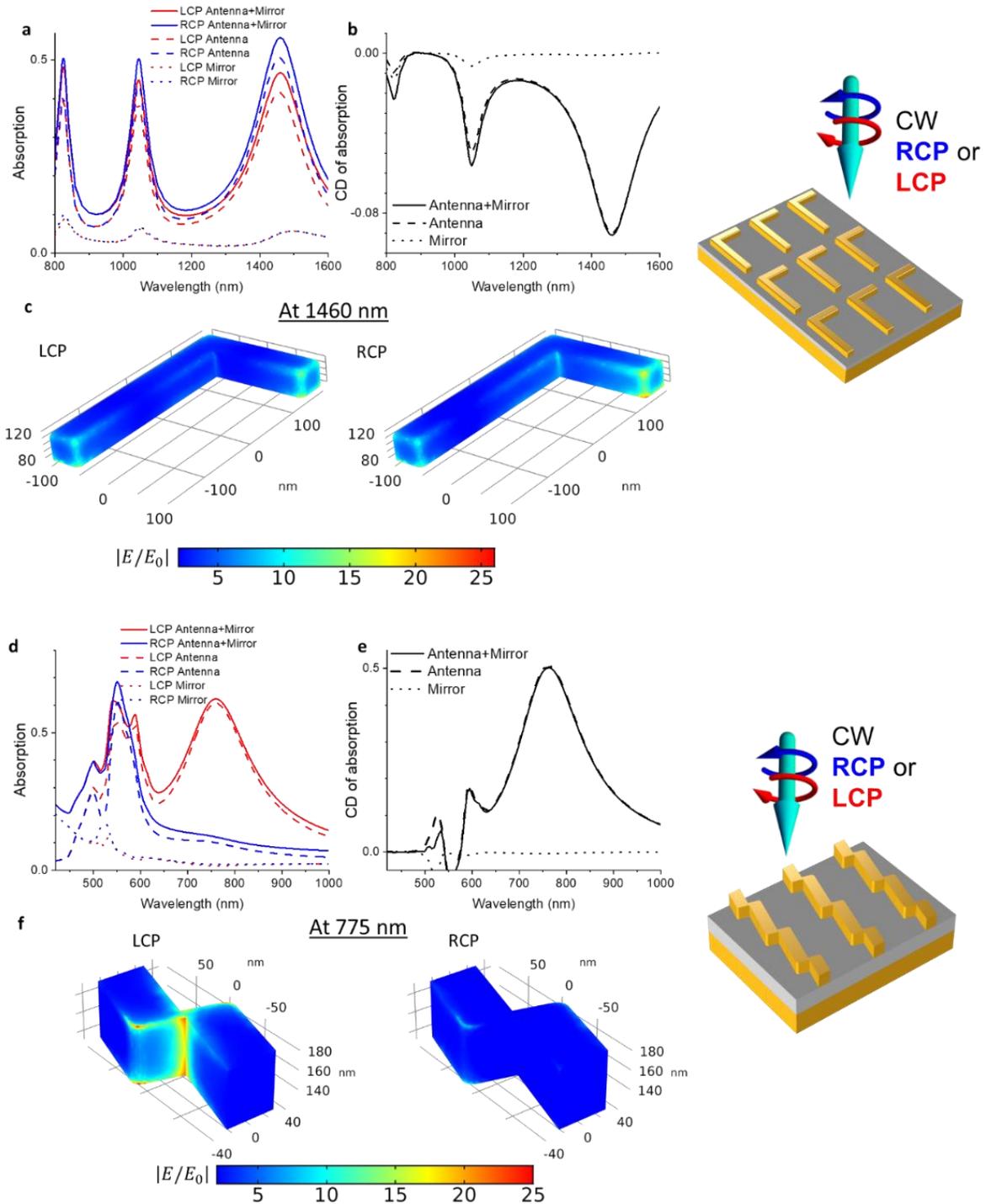

Figure 2: Linear responses of chiral metastructures. (a-c) for the chiral L-shaped metastructure, and (d-f) for the chiral wire, with several unit cells periodic schematics shown on top panels. (a) and (d) are the absorption spectra for the antenna (dashed lines), for the mirror (dotted lines), and for the antenna-mirror complex (solid lines), all for LCP (red) and RCP (blue) illumination. (b) and (e) are the circular dichroism (CD) of absorption for the antenna (dashed line), for the mirror (dotted line), and for the antenna-mirror complex (solid line). (c) and (f) are the spatial distribution of the electric fields when the structure is excited at the plasmon of maximum CD.

First, we study the optical response of each metastructure in the CW regime only. Figures 2 and 3 show the electromagnetic responses of the systems in such regime, for the chiral and the achiral metastructures, respectively. For the chiral L-shaped and wire, Figs. 2a and 2d show the simulated absorption $A_\alpha$ spectra, and 2b and 2e show the corresponding CD, individually for the Au antenna, the Au mirror and the combined system. Figures S16a and S16b show the corresponding g-factors. For our first chiral metastructure, the L-shaped, we can see that the absorption spectra show three main peaks, the most important being at 1460 nm, where both the antenna's CD and the g-factor are maximum (~|0.9| in Fig. 2b and |0.2| in Fig. S16a, respectively). At this pumping wavelength the systems show the largest absorption for RCP irradiation, given that the geometry is right-handed, occurring the opposite when the L-shape is left-handed (meaning reflected in the plane along the L's long axis, not shown). Most of the absorption occurs at the antenna (~90%) rather than the mirror (~10%), with about $A_{RCP} \approx 0.56$ and $A_{LCP} \approx 0.47$ for the whole system, while the rest is reflected (not shown). The plasmon hot-spots are located at the short arm of the L-shape, specifically at its 4 corners, as shown in Fig. 2c (see Fig. S12 for the plasmon full distribution), being slightly larger in magnitude for RCP than LCP because of the geometry. The second chiral structure on the other hand, the chiral wire, is a different case since it acts more as a chiral filter at its main peak, allowing mostly $A_{LCP}$ while $A_{RCP}$ being significantly smaller. One can see in Fig. 2d that for the main plasmon at 775 nm, $A_{LCP} \approx 0.6$ while $A_{RCP} \approx 0.1$, each mostly contributed by the antenna absorption (98% for LCP, 86% for RCP). Correspondingly, the antenna's CD (~0.5 in Fig. 2e) and the g-factor (~|1.4| in Fig. S16b) are maximum at 775 nm, providing a better chiral metastructure. The plasmon for LCP irradiation localized very strongly (~3x RCP) at the thinner part of the wire across the whole vertical dimension, and for RCP only at the outer corners of the wire, as shown in Figs. 2f and S13. Even though we choose a fixed set of materials and geometry dimensions, in general one can modify these plasmonics responses by playing with the matrix material, the gold's geometry and polarization of light. We have previously studied these structures in other contexts, and we have shown how their electromagnetic response can be largely controlled, both the L-shape[39] and the chiral wire[52,53].

Figure 3 shows the electromagnetic responses of the achiral metastructures under study, schematics at the top row of each panel, $A_\alpha$ on the middle row, and main plasmon distribution at the bottom row (3a and 3e for the disk, 3b and 3f for the rod, 3c and 3g for the rectangular prism, and 3d and 3h for the achiral wire). Each system presents a main plasmon in the visible region, with the rectangular prism and the rod presenting two, a transversal (at lower wavelength) and a longitudinal (at higher wavelength); we will focus only on the latter. Each metastructure shows a high degree of absorption at the plasmon resonance (disk $A_\alpha$ ~0.8 for 601nm in Fig. 3a, rod $A_\alpha$ ~0.5 for 676nm in Fig. 3b, rectangular prism $A_\alpha$ ~0.6 for 682nm in Fig. 3c, and achiral wire $A_\alpha$ ~0.55 for 560 in in Fig. 3d), for either LCP or RCP, leading to true zeros of the corresponding CDs and g-factors, as physically expected (see Figs. S14-15 and S16 for CD and g-factors, respectively). Again, the larger contribution to the absorption comes from the antenna rather than the mirror. As for the spatial distribution the electromagnetic fields, whereas the disk and the achiral wire do not show any near-field chirality (Figs. 3e, 3h, S10, and S11), the rectangular prism and the rod they do show some: opposite corners at the top face are excited when illuminating with LCP (Figs. 3f and 3g, S8 and S9), while the opposite corners are excited when illuminating with

RCP (Figs. S8 and S9). This effect was recently observed in experiments, which lead to the chiral growth of selective prism corners[60]. This static-induced chirality we see in our CW simulations, depends on the symmetry of the metastructure lattice, which arises due to the quasi-2D reduced dimensionality ('quasi' since they are truly 3D objects). We have previously addressed the symmetries and operations describing many substrate experiments with unidirectional light illumination, in a '2D vs 3D' chiral competition (See last sections of Refs. [4,61]). In general, we can separate our systems in two groups depending on their symmetry. The disks array can be seen as a square lattice, the rectangular prisms and the rods arrays can be seen as rectangular lattices, and the achiral wire can be seen just a periodic 1D lattice. As such the disk metastructure possesses a number 11 *p4m* symmetry group (has translations, rotations of 90° and 180°, reflections, glide reflections, and inversion). The rectangular prism and the rod metastructures possess a number 6 *pmm* symmetry group with no inversion (has translations, rotations of 180°, and reflections). If the metastructure belongs to the first class, no chirality arises; if it belongs to the second class, near-field chirality arises. We term this effect "imprinting of chirality".

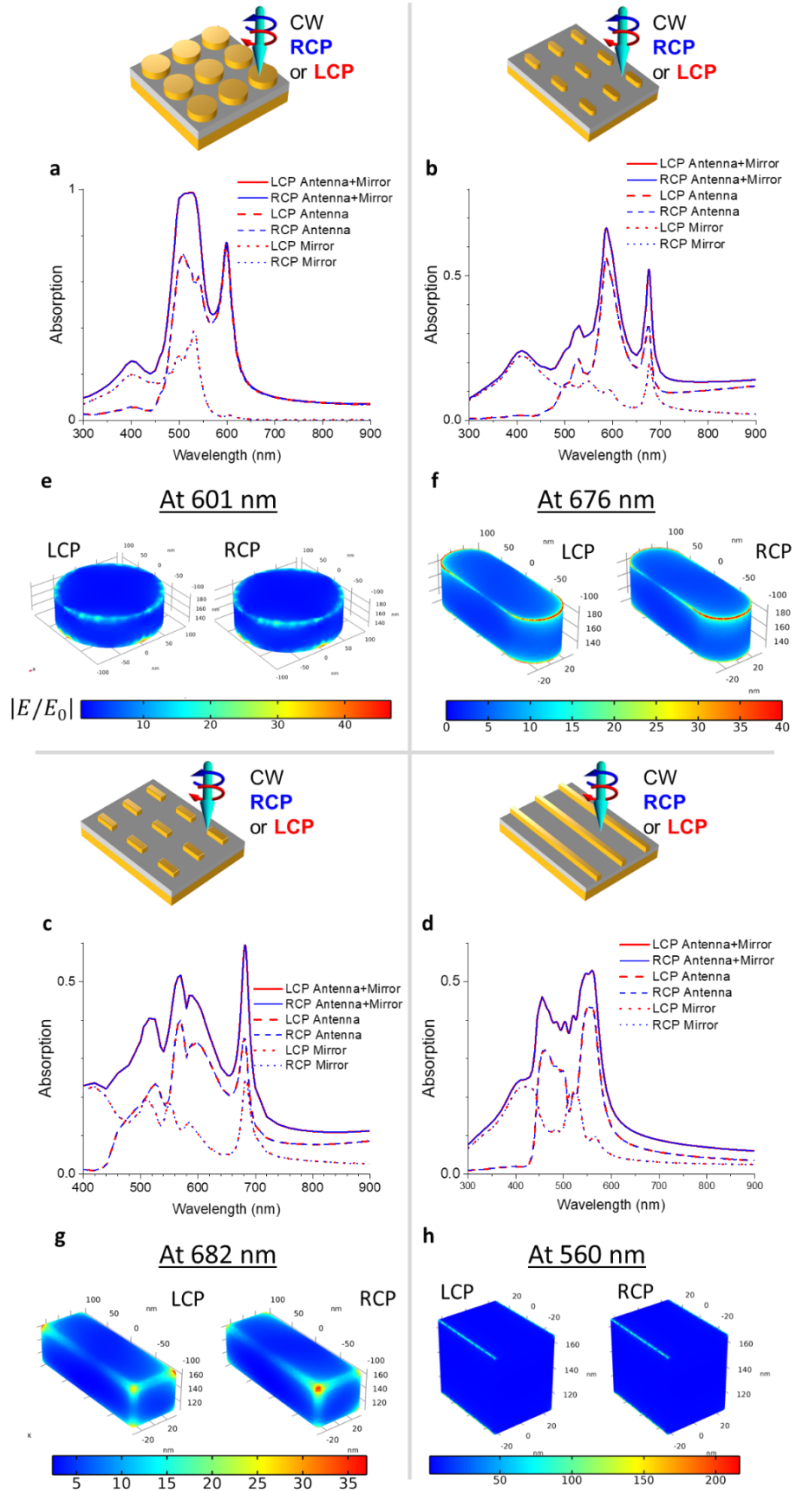

**Figure 3: Linear responses of achiral metastructures.** Several unit cells for each structure are schematized on top panels: (a-d) Absorption spectra for antenna (dashed lines), for mirror (dotted lines), and for antenna-mirror complex (solid lines), all for LCP (red) and RCP (blue) illumination; (e-h) are the spatial distribution of the electric fields when each structure is excited at the main plasmon resonance (For full electromagnetic maps, see Figs. S7-S10). In all cases, the maximum magnitudes of the spectra for the circular dichroism (CD) are $\leq 10^{-14}$ and g-factors are $\leq 10^{-13}$ (not shown, see Sup. Material Figs. S14 and S15).

## IV. Imprinting of T-dependent Chirality in Chiral Metastructures

Given that several of the calculations in the field are performed within the CW regime, which assumes a constant illumination on the metastructure yielding steady state patterns, but more realistically, experiments are perform with by pumping an ultrafast laser and proving the system's response, which might lead to think that the laser effects could be just temporary. So, an obvious question is whether this chirality imprinting fades out over time, or if it remains. If so, at what time scales does it occur? We address now in what follows. Figures 4 and 5 show the dynamical CPL pump-probe simulations within the 2T-model only for chiral and achiral metastructures, respectively. We simulate a typical pump-probe experimental scenario, where an LCP pump irradiation is set to last 100fs via a Gaussian-time profile pulse, exciting the metastructures, for which we record the chiral response via a CPL probe.

We first focus on the chiral structures, shown in Fig. 4. for the chiral wire (a-c) and for the L-shaped (d-f). Figures b and e show for the chiral wire and the L-shaped, respectively, the averaged pumped temperatures for either model after a 100fs gaussian pulse (in magenta, right axis), this is before applying the probe beam. The set of computational data shows that the temporal chiroptical responses are sensitive to geometry, revealing the formation of chiral hot-spots (Figs. S20-S23). The most intriguing effect in these complex patterns is the prediction of the ability to imprint temporal chirality on any solid using CPL and anisotropic metastructure.

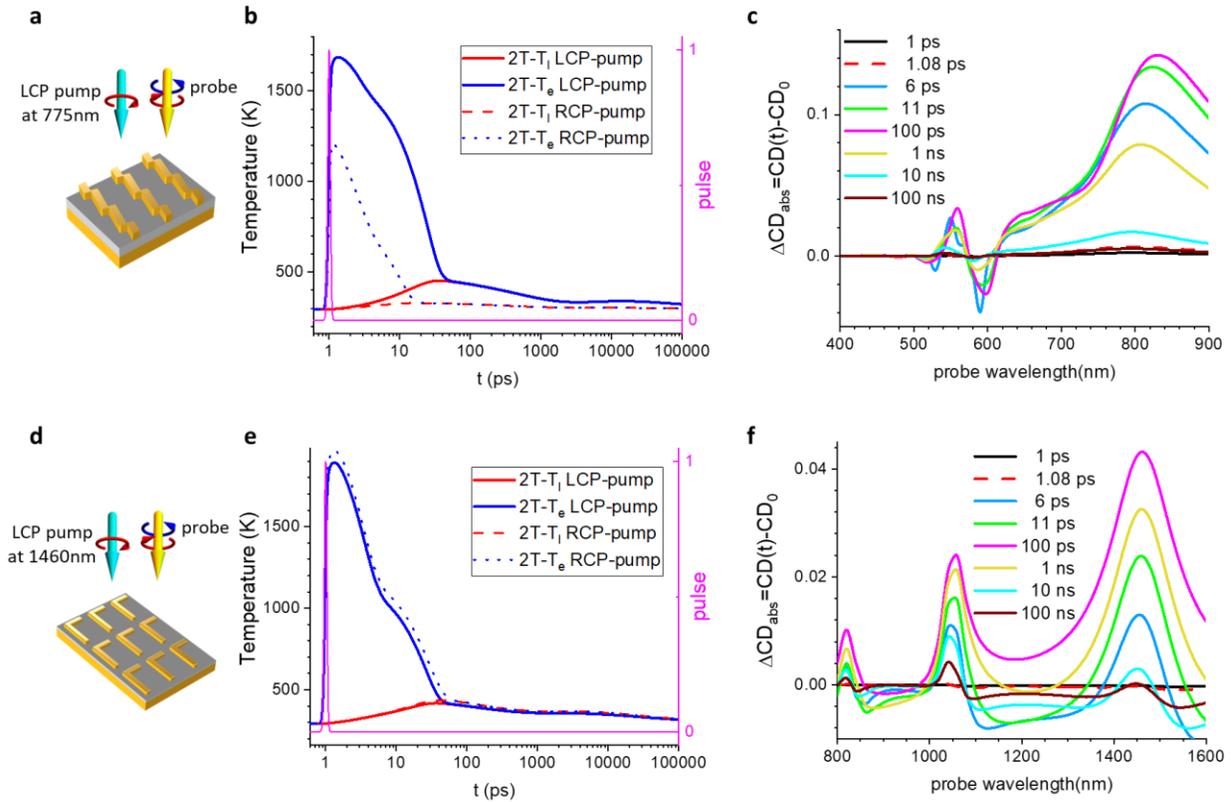

**Figure 4: Dynamical CPL pump-probe simulation for chiral metastructures**. CPL photoexcitation and temporal-thermal responses for the chiral wire (a-c) and for the L-shaped metastructures (d-f), both setups for LCP pump and LCP/RCP probes as schematized in (a) and (d), respectively. (b) and (e) are the volumetric temperature averages over the entire

structure for the 2T (red for $T_l$, blue for $T_e$) model, for either pumping LCP (solid lines) or RCP (dashed lines), after the 100 fs pulsed illumination (pink line, right axis). (c) and (f) are the CD difference $\Delta CD_{abs} = CD(t) - CD_0$, this is between the CD at any specific time $t$ and the CD at with CW for the 2T model. Individual $CD(t)$ curves are provided in see Sup. Material, Fig. S16, whereas $\Delta CD_{abs}$ for the 1T model is shown in Fig. S19. Both metastructures were simulated with an intensity $I_0 = 10^{11}$ W/cm$^2$, and both pumped at the plasmon frequency described in the CW regime (see Fig. 2) with a CPL beam $\vec{E} // -k_z$.

In order to understand these interesting, nontrivial behaviors and to shine light on how they can be tracked experimentally, we apply the CPL probe for the pumped structures. Individual CPL probe CD spectra were calculated for several delay times (see Fig. S18), from which we then subtract the CD spectrum calculated for the CW regime (which we call $CD_0$), meaning $\Delta CD_{abs}(t) = CD(t) - CD_0$, obtaining a quantitative measurement of the departure from the static case. The $\Delta CD_{abs}(t)$ within the 2T model are shown in Figs. 4c for the chiral wire, and 4f for the L-shaped (see Fig. S19 for comparison with the 1T model). When looking at individual CD spectra (Fig. S18), in general we can see that the levels of chirality for either model are different due to their inherent geometry: whereas the chiral wire shows a maximum $|CD_0| \approx 0.5$, which red-shifts and increases up to $|CD(t=10ps)| \approx 0.7$, the L-shaped system shows a maximum $|CD_0| \approx 0.04$, which red-shifts and decreases down to vanishing $|CD(t)|$ (~100 ps). Moreover, the fact that the chirality can be increased or decreased is a signaling of the chirality imprint we proposed earlier, occurring at different time scales. The 2T model shows consistent dynamics altogether (Figs. 4c and 4f). The departure from the CW case occurs at larger time scales, and not immediately around the pulse. We can see that $\Delta CD_{abs}(t)$ becomes finite a few ps after the pulse obtaining it maximum about 100ps and then decreasing until eventually recovering the CW response, which of course could be controlled with the beam's intensity (Fig. S17).

## V. Imprinting of T-dependent Chirality in Achiral Metastructures

We now focus on the pump-probe dynamics of achiral structures, shown in Fig. 5, with the rod array in panels (a,c,e), the rectangular prism in panels (b,d,f), and both disk and achiral wire in panels g-h. Figures 5c and d show the volumetric average temperatures after LCP and RCP pump, leading to complex time dependences with $\tau_{e-ph}$ ~6 ps (rod) and ~8 ps (rectangular prism). For the other structures and comparison with the 1T model see Fig. S24 and S25, and for intensity dependence see Fig. S26. Spatial distributions of the temperatures $T_l$ and $T_e$, respectively, are shown in Figs. S27 and S28 for the disk, S29 and S30 for the rod, S31 and S32 for the rectangular prism, and S33 and S34 for the achiral wire. We see that no pumped chiral pattern appears neither for the disk nor the achiral wire, as expected. The rod and the rectangular prism on the other hand, show significant chirality imprinting at opposite corners of their rectangular geometries. In general, the chiral thermal patterns appear first on the $T_e$, which is later transferred to the phonons in a few ps, which in turns, is notable in the probe signatures now studied.

On the following pump-probe analysis for the 2T model, one must note that given that we are using the symmetric mesh approach our metastructures show a true zero upon CPL illumination, $CD_0 \approx 10^{-14}$ (Fig. S14 and S15), such that $\Delta CD_{abs}(t) \equiv CD_{abs}(t)$, which is why we will focus only on $CD(t)$. The 2T model shows a clear dynamic at similar time scales. Neglecting the disk and the achiral wire, where CD values are several order of magnitude smaller than the rod and rectangular prism, their maximum CD values are $|CD(t=5ps)| \approx 0.002$ and $|CD(t=5ps)| \approx 0.005$, respectively, showing also some finite chirality at ~10 ps, much latter than the pulse set at $t_0$. This suggests that whereas in spherical NPs the hot electrons are fully thermalized after ~10 ps, in the metastructures here studied, the thermalization occurs at much larger time scales, depending on the intensity. Here lies a key issue of our results: whereas the pump does not show any chirality at all (LCP=RCP in Figs, S24b,e,h,k), the probe setup is necessary to catch the chirality imprinting in the systems, at some level with the 1T-model and more reliably with the 2T-model. We can also see that by examining the thermal dissipation after the electron–phonon coupling regime, all of our metastructures show large heat gradients between core and surfaces mostly at>10ps timescale.

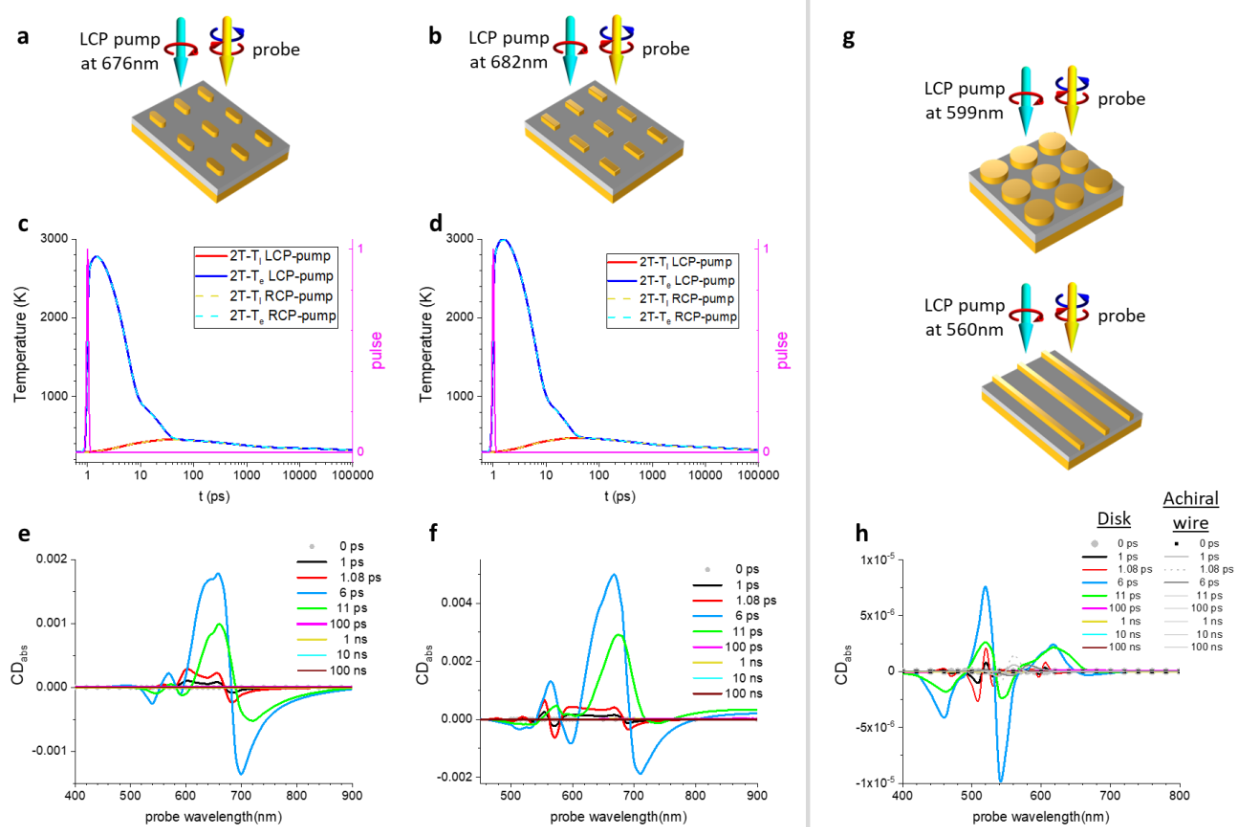

Figure 5: **Dynamical CPL pump-probe simulation for achiral metastructures**. CPL photoexcitation and temporal-thermal responses for the rods (a,c,e), for the rectangular prisms (b,d,f), and for both the disk and the achiral wire (g-h) metastructures, all for LCP pump and LCP/RCP probes. (c) and (d) are the volumetric temperature averages over the entire structure for the 2T (red/yellow for $T_l$, blue/cyan for $T_e$) model, for either pumping LCP (solid lines) or RCP (dashed lines), after the 100 fs pulsed illumination (pink line, right axis). (e,f,h) are individual $CD(t)$ curves for the times there specified, obtained via the 2T model. Note that panel (h) shows curves for two systems, the disk and the achiral wire; while for the first the color curves are same as in other panels, for the second we show the curves in different shapes of gray only, as they are nearly negligible when compared to the other systems. For all panels we also include the CW CD, $CD_0$, which is inherently zero, as shown in Fig. S15. All structures were simulated with an intensity $I_0 = 10^{11}$ W/cm², and pumped at the plasmon frequency described in the CW regime (see Fig. 3) with a CPL beam $\vec{E} // -\vec{k}_z$.

## VI. Discussion – Chiral Light Creates Chirality, But There Are Limitations

In Fig. 6 we see the spatial distribution maps of the photo-induced CD with the pump-probe setup, as defined in Eq. 14, for the chiral wire (6a-b), for the L-shaped (6c-d), and for the rectangular prism (6e-f). See Fig. S35 for the other metastructures. We see that zones of different chirality can be excited with either CPL. Along different vertical cuts, the chirality is stronger at the antenna and changes via the distinct excited multipoles, vanishing throughout the glass media. For the achiral metastructures, one can see that the rods and the rectangular prisms show a non-vanishing chirality, as spectrally anticipated in Figs. 5e and 5f. The spatial distribution concentrates on the antenna's top as a quadrupole and at the bottom as an octupole, because of its connection to the Au mirror.

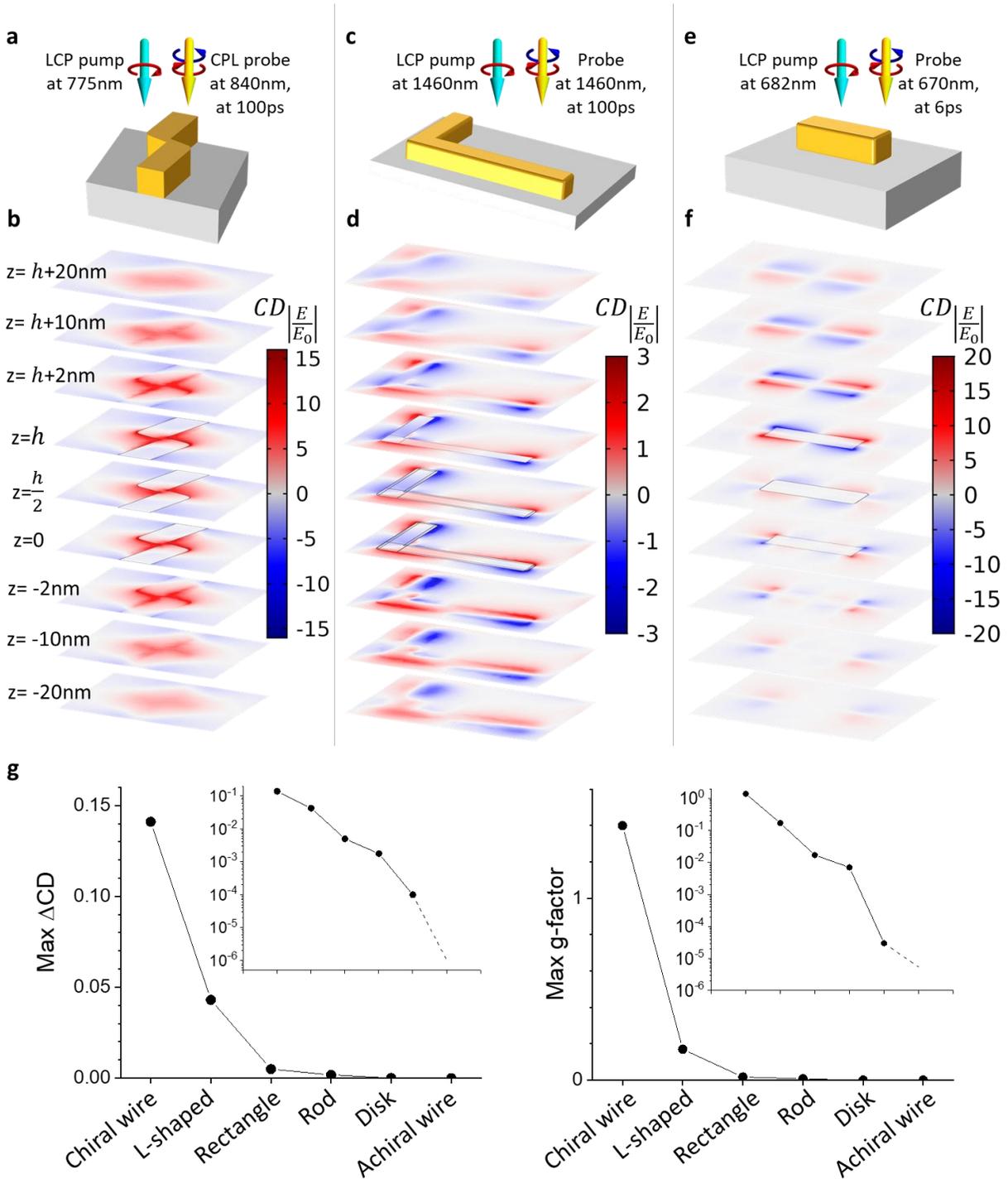

Figure 6: Electronic CD for the pump-probe simulations. CD of the electric field distributions for the chiral wire (a,b), for the L-shaped (c,d), and for the rectangular prisms (e,f), at the times t=100ps, t=100ps and t=6ps, respectively, this is when each structure shows a maximum absorption CD in Figs. 4c, 4f and 5h respectively, within the 2T-model. For the rest of the structures see Fig. S35. For clarity, at the top row we show the correct orientation of the unit cell, for the plane cuts shown at the bottom row. At the bottom row (panels b,d,f), each plane is a two-dimensional cut as indicated on the left of panel b, where $h$ is the height of each antenna, as indicated in Figs. S6 and S7 for each structure, such that $z = 0$ corresponds to the bottom of each Au antenna. Panel (g) shows the maximum ΔCD and g-factors induced upon irradiation.

The difference in CD intensity upon irradiation, driven by the induced thermal modifications to the dielectric function, yielding diverse quasi-2D multipolar distributions, is the phenomena we termed as chirality imprinting, which only lasts a few ps after irradiation (can be controlled with the incident intensity). In order to qualitatively classify such an effect in Fig. 6g we show the maxima CD and g-factors when probing the metastructures. We can identify three regimes: chiral (chiral wire and L-shaped), weak chiral (rectangular prism and rod) and achiral (disk and achiral wire). Our optical dynamical temporal evolution simulations could be tested via ultrafast transient absorption spectroscopy[62] in the far field or local photothermal microscopy.[63,64] Local electromagnetic chirality can also be visualized by the near-field microscopies[65,66] and AFM[67,68].

We should also comment on how this effect should be observable in other materials. Similar metastructures as the ones here described are also very common in the literature or should be in experimental reach, since they can be built in periodic metastructures and show plasmons in the visible/near-IR. As such, a similar approach to ours can be carried out. The procedure starts by describing the dielectric function with a Drude-like model, as the one in Eq. 4. We have fitted Eq. 4 to several experimentally available data for silver (Ag)[69], copper (Cu)[70], and titanium nitride (TiN)[71], and found suitable values for $\varepsilon_{b,\text{Drude}}$, $\omega_{p0}$, and $\gamma_{D0}$. The procedure continues by finding suitable thermal quantities for these materials, which are largely available in the field. Table 1 shows all these parameters of these massively studied plasmonics nanomaterials. Hence, our approach could be straightforwardly applied to these and other materials in order to study chirality imprinting patterns in them.

**Table 1.** Parameter comparison for the current plasmonic materials of interest: gold (Au), silver (Ag), copper (Cu) and titanium nitride (TiN).

|  | Au | Ag | Cu | TiN |
|---|---|---|---|---|
| $\varepsilon_{b,\text{Drude}}$ | 9.07 | 3.15 | 5.5 | 8.765 |
| $\omega_{p0}$ [eV] | 9.1 | 8.85 | 8.184 | 6.941 |
| $\gamma_{D0}$ [eV] | 0.073 | 0.037 | 0.032 | 0.297 |
| Density [Kg/m³] | 19300 | 10490 | 8960 | 5400 |
| Melting temperature [C] | 1063 | 961 | 1084 | 2930 |
| Thermal conductivity, $\kappa$ [W/(m*K)] | 315 | 429 | 398 | 29 |
| Heat capacity, $C_P$ [J/(Kg K)] | 126 | 233 | 386 | 784 |

The choice of these materials in particular, is because of great interest to the community as alternatives for gold. Silver is known to be the best plasmonic material with lowest ohmic losses at optical frequencies and in general producing stronger and sharper resonances that gold, which widespread implementation has been limited due to technological challenges in thin fabrication techniques[72]. Going beyond Au and Ag[73], candidates such as Cu and TiN show great promising. Cu-metastructures show an interband transition energy level of 2.15 eV, hence having the capability of interband excitation within the visible range, although oxidation of its surfaces is still an issue[74]. On the other hand, TiN has proved to be a good plasmonic material in the vis/NIR region, and because is a good refractory material, has a high melting point and high chemical stability in harsh environments[75,76]. Mostly, disk-shaped TiN metastructures have been proposed[77,78] and some already grown[79,80].

Finally, me must note that the term chirality imprinting has been addressed before in other contexts. In quantitative chemosensing, Bentley et al[81] demonstrated that a stereodynamic scaffold probe carrying complementary boronic acid and urea units gets instantaneous chirality imprinting in the UV, allowing to readout its corresponding target compound. Just in the past year, 2 groups have reported 'chirality imprinting' results. First, chiral superpositions of atomic states were proved to be excitable with tailored light fields both in the weak-field and strong-field regimes, using realistic laser parameters, where the chirality was imprinted on the photoelectron wave packet created by strong-field ionization, for both hydrogen and sodium atoms[82]. Lastly, a new method called circular depolarization spectroscopy was introduced to study photo-imprinting of chirality in materials made of photo-isomerizable groups, meaning that the materials reorientate their molecules in a cholesteric organization in a particular handedness[83].

## VII. Conclusions

Using realistic models of achiral and chiral metastructures, we demonstrate the chirality imprinting in anisotropic metastructures, both achiral and chiral. The mechanism of chiral imprinting is owing to an interplay between the external CPL and the strong local fields in a metastructure. The choice of geometry is crucial for our effect to be strong and measurable. Isotropic and high symmetry metastructures, like a single disk or nanowire array, do not show any imprinted chirality for a fundamental symmetry reason. This demonstrates the importance of modeling and theoretical understanding. The observed temporal chiral effects are multiscale – they are due to a combination of electric-field distributions (at shorter times!) and thermal distributions affecting the optical response (at longer times!). Our findings and models could be useful in a plethora of optical and biosensing applications, where tailor-made thermal responses are needed at specific time scales.

ASSOCIATED CONTENT

**Supporting Information.**

Supporting Information is available free of charge from the publisher's website.

AUTHOR INFORMATION

**Corresponding Authors**
O.A.-O., Z.W., G.M., and A.O.G

**Notes**

The authors declare no competing financial interest.

**Acknowledgments**

Z.M.W. acknowledges the National Key Research and Development Program of China (2019YFB2203400) and the "111 Project" (B20030). O.A.-O., G.M., and A.O.G acknowledge the generous support from the United States-Israel Binational Science Foundation (BSF), grant number 2018050. O.A.-O. and A.O.G are supported by the Nanoscale & Quantum Phenomena Institute (NQPI), the Quantitative Biology Institute (QPI), and the OUCR/Baker fund at Ohio University. L.V.B. acknowledges the support from the Spanish Ministerio de Ciencia e Innovación through a Ramón y Cajal fellowship and under projects PID2020-118282RA-I00 and TED2021-130038A-I00, and the National Natural Science Foundation of China (Project No. 22250610200). O.A.-O. acknowledges fruitful discussions with Dr. Sven H. C. Askes and Dr. Larousse Khosravi Khorashad.

## ORCID

Oscar Ávalos-Ovando: 0000-0003-3572-7675

Veronica A. Bahamondes Lorca: 0000-0002-0488-2472

Lucas V. Besteiro: 0000-0001-7356-7719

Artur Movsesyan: 0000-0002-5425-7747

Zhiming Wang: 0000-0003-4171-1821

Gil Markovich: 0000-0002-4047-189X

Alexander O. Govorov: 0000-0003-1316-6758

Supporting Information for

# Universal imprinting of chirality with chiral light by employing plasmonic metastructures

Oscar Ávalos-Ovando[1,2*], Veronica A. Bahamondes Lorca[3,4], Lucas V. Besteiro[5], Artur Movsesyan[1,6], Zhiming Wang[6], Gil Markovich[7*], and Alexander O. Govorov[1,2*]

[1] Department of Physics and Astronomy, Ohio University, Athens, Ohio 45701, United States

[2] Nanoscale and Quantum Phenomena Institute, Ohio University, Athens, Ohio 45701, United States

[3] Edison Biotechnology Institute, Ohio University, Athens 45701, OH

[4] Departamento de Tecnología médica, Facultad de Medicina, Universidad de Chile, Santiago, Chile

[5] CINBIO, Universidade de Vigo, 36310 Vigo, Spain

[6] Institute of Fundamental and Frontier Sciences, University of Electronic Science and Technology of China, Chengdu 610054, China

[7] School of Chemistry, Raymond and Beverly Sackler Faculty of Exact Sciences, Tel Aviv University, Tel Aviv, 6997801 Israel

* Corresponding authors: oa322219@ohio.edu, gilmar@tauex.tau.ac.il, govorov@ohio.edu

# I. Temperature dependent quantities for the Drude model, and for the 1T-and 2T-models

**Table S1.** Drude-Sommerfeld parameters of the metal gold-like model used in our nanostructures.

| Drude parameter | value | unit | Reference |
|---|---|---|---|
| $\varepsilon_{b,Drude}$ | 9.07 | [1] | 1 |
| $\omega_{p0}$ | $1.327 \times 10^{16}$ | [rad/s] | 1 |
| $\gamma_{D0}$ | $1.09 \times 10^{14}$ | [rad/s] | 1 |
| $\alpha_V$ | $4.5 \times 10^{-5}$ | [1/K] | 2 |
| $\gamma_{e-ph}$ | $4.56 \times 10^{13}$ | [rad/s] | 3 |
| $\beta$ | $3.3 \times 10^{-3}$ | [1/K] | 4,5 |

**Table S2.** Thermal and optical parameters of the materials used in our nanoparticles and metastructures.

| Material | Thermal conductivity, $\kappa \left[ \text{W m}^{-1} \text{ K}^{-1} \right]$ | Mass density, $\rho \left[ \text{Kg m}^{-3} \right]$ | Heat capacity at constant pressure, $C_p \left[ \text{J Kg}^{-1} \text{ K}^{-1} \right]$ | Dielectric constant |
|---|---|---|---|---|
| Water | 0.6 | 1000 | 4181 | 1.8 |
| Glass | 1.4 | 2650 | 840 | 2.13 |
| Au Lattice | Fig. S2b | Fig S2c | Fig. S2a | Eq. 8 (for CW) Eq. 13 (for pump-probe) |
| Au Electrons | Fig. S2b | - | Fig. S2a | - |

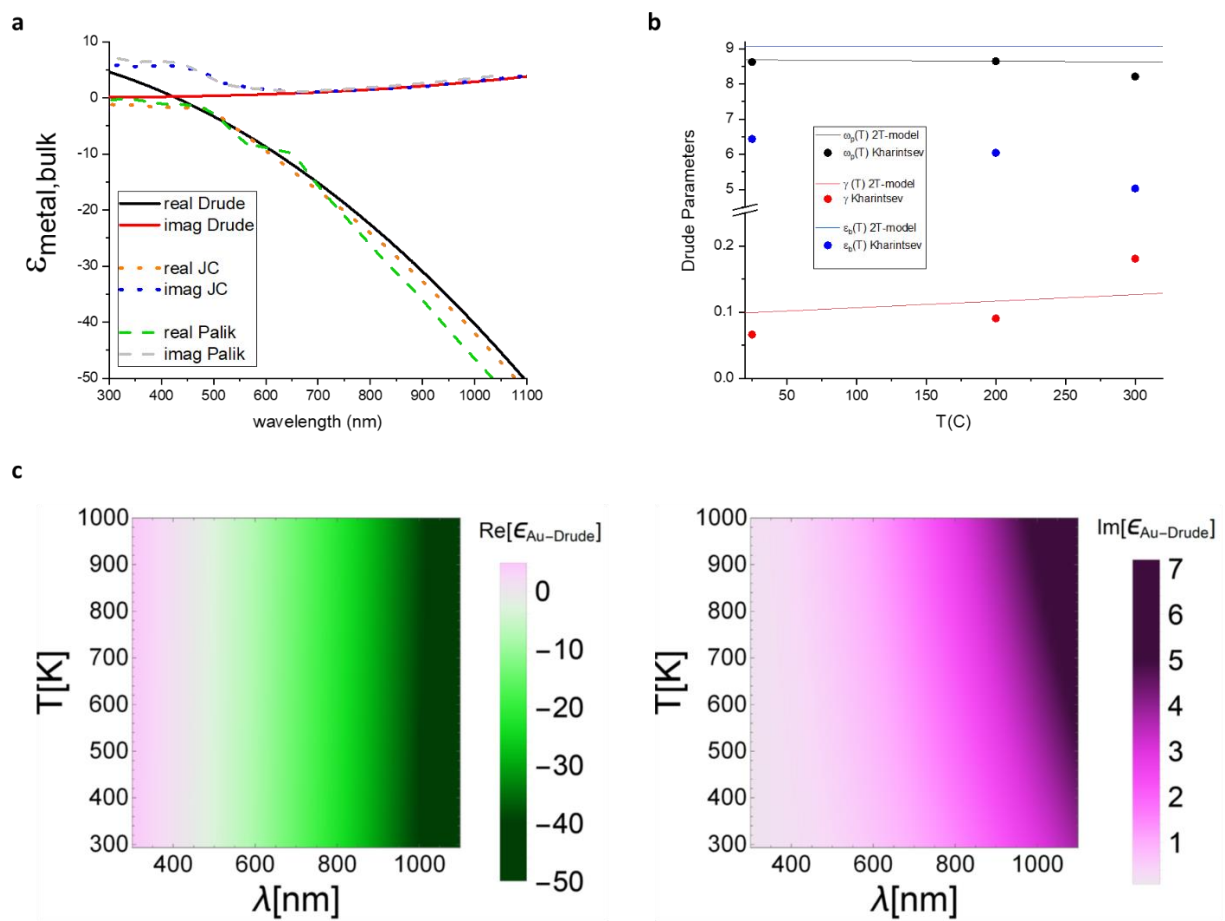

**Figure S1:** (a) Gold dielectric functions, both the theoretical Drude used throughout this work and two of the experimental vastly used in the literature[6,7]. (b) Temperature-dependent Au-Drude parameters used in our work (lines) in Eq. 5 compared to recent experimentally measured values[8] (symbols). (c) Real and imaginary parts of the temperature-dependent Au-Drude dielectric functions from Eq. 5.

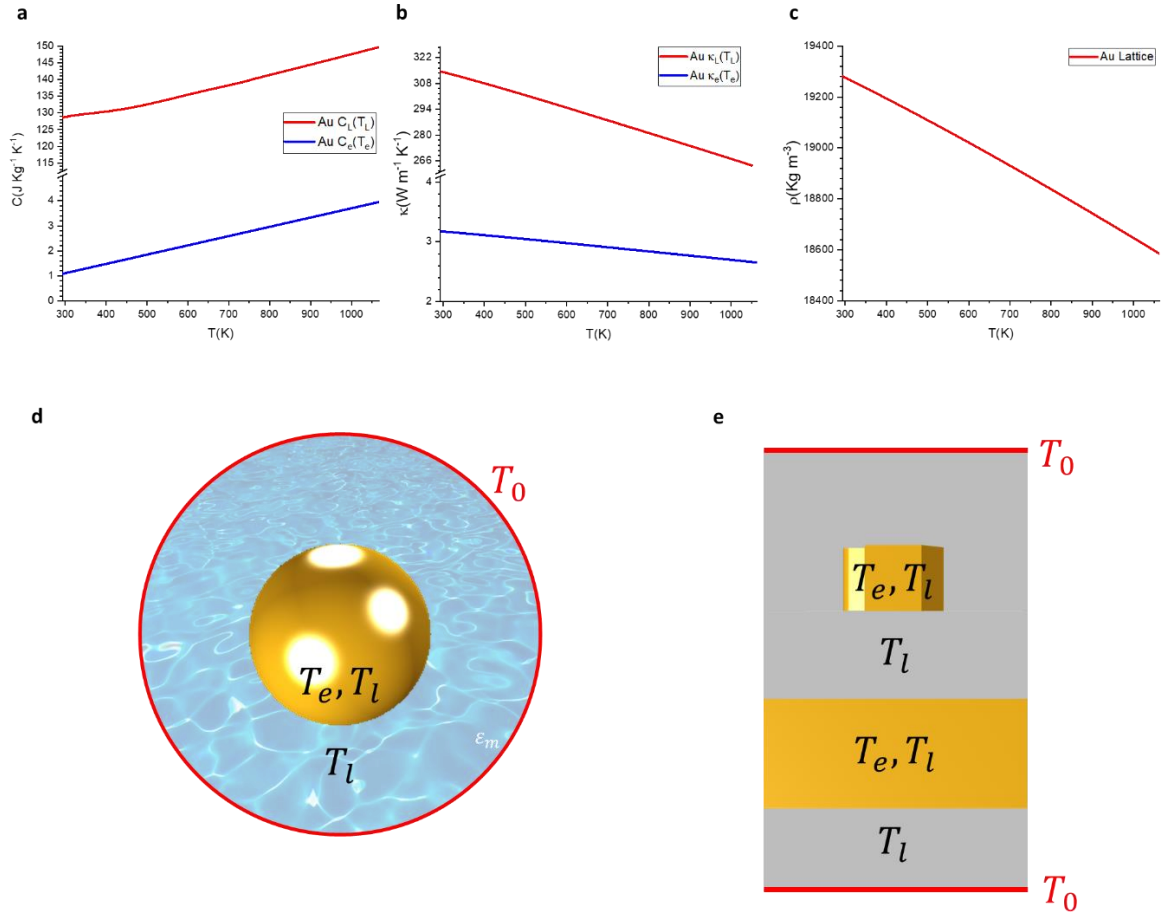

**Figure S2:** Gold temperature-dependent quantities and thermal setups used throughout this work: (a) Heat capacity at constant pressure, (b) thermal conductivity, and (c) density. Panels (d) and (e) show schematics (not-to-scale) of the boundary conditions for the materials (gold in golden color, water in light blue, glass in gray) and their temperatures considered in our work: the outermost boundaries are always kept at $T_0 = 293.15$ K and inside both lattice ($T_l$) and electron ($T_e$) temperatures are allowed to interact in the 2T model, or just $T_l$ in the 1T model. Also, $T_l$ exist everywhere, whereas $T_e$ exist only in the metal, which in our case is gold. The schematics in panel (e) are repeated infinitely along the plane and the out of plane directions here shown.

# II. Validation of the 1T- and 2T-models with a gold nanoparticle

Here we first validate our 1T- and 2T-models. We study a simple gold nanostructure, i.e. a spherical gold NPs, surrounded by water, in order to gain perspective on how both heat dissipation models behave. The differences between the 1T and the 2T model in simple gold spherical NPs, of two typical real sizes (diameters 20 and 100 nm) are shown in Figs. S3 (d=20 nm NP) and S4 (d=100 nm NP), when illuminated with a linearly polarized beam. In both cases, the NPs have similar dipolar electric field enhancements upon illumination (Figs. S3a and S4a), and opposite pole locations of the maximum total power dissipation densities (Figs. S3b and S4b). After the 100 fs gaussian illumination pulse, the thermal dynamics of the 2T model begin: the electron temperature increases after an exciting pulse, then relaxes and transfers energy to the lattice, exciting the lattice phonons, hence increasing the lattice temperature. Finally, when all the energy is transferred from electrons to phonons, the systems reach an equilibrium temperature with the surrounding media (see Figs. S3c and S4c). For 1T model on the other hand, the lattice temperature just increases up to a maximum immediately after the pulse, and then monotonously decreases. All the temperatures, $T_l$ and $T_e$ from the 2T model and $T_l$ from the 1T model, meet at large times, about 100 ps for the 20nm NP and about 3 ns for the 100nm NP. As for the temperature spatial distribution, we can see that the small 20nm NP is initially equally heated at either the top or the bottom poles for either model (Fig. S3c, right panel), and then the temperatures decay equally over time, probably due to the fast heat transfer throughout the small volume. The larger 100nm NP shows differences for the north and the south poles for either model: the larger lattice temperature is located at the south pole, the one closer to the incoming beam, and then a time passes the heat is distributed through phonons until both poles equalize ~11 ps for the 1T model and ~150 ps for the 2T model (Fig. S4c, right panel). Additionally, steady-state longitudinal temperature profiles along the NP's axes (Figs. S3d and S4d), show excellent agreement between the lattice temperatures of our 1T and 2T models, with only one marked difference: the 2T models show that the NP heats up more in the middle than in the edges, whereas in the 1T model the lattice temperature is constant throughout the entire NP's profile. As for the electronic temperature, whereas along the E-field polarization directions (x and y) the temperature is larger in the NP's center and decreases towards the edges, along the irradiation direction (+z) the electrons are hotter opposite to the point of the incoming wave (top pole) for the small 20 nm NP, but hotter on the closest point of the incoming wave (bottom pole) for the large 100nm NP. We attributed this behavior to the fact that in the largest NP, the Te and Tl are closer together (Fig. S4d), so energy exchange between electrons and phonons should occur faster.

We can estimate the electron-phonon couplings ($\tau_{e-ph}$) within the 2T model, defined as the time that the maximum $T_e$ decreases up to $T_e/e$, where we estimate them to be 0.96 ps for the 20 nm NP and 0.83 ps for the 100 nm NP, in good agreement with the vast literature,[9–11] whereas our phonon-phonon relaxation times are in the order of several tens of ps, in good agreement with literature as well.[12] Note that these $\tau_{e-ph}$ in principle change with the irradiation intensity as shown in Fig. S5. In our following calculations we will use artificially large intensities, since we want to identify significant changes in the temperature dependent Drude model of Eqs. 7 and 8. With these results we validate our model and continue to more complex structures.

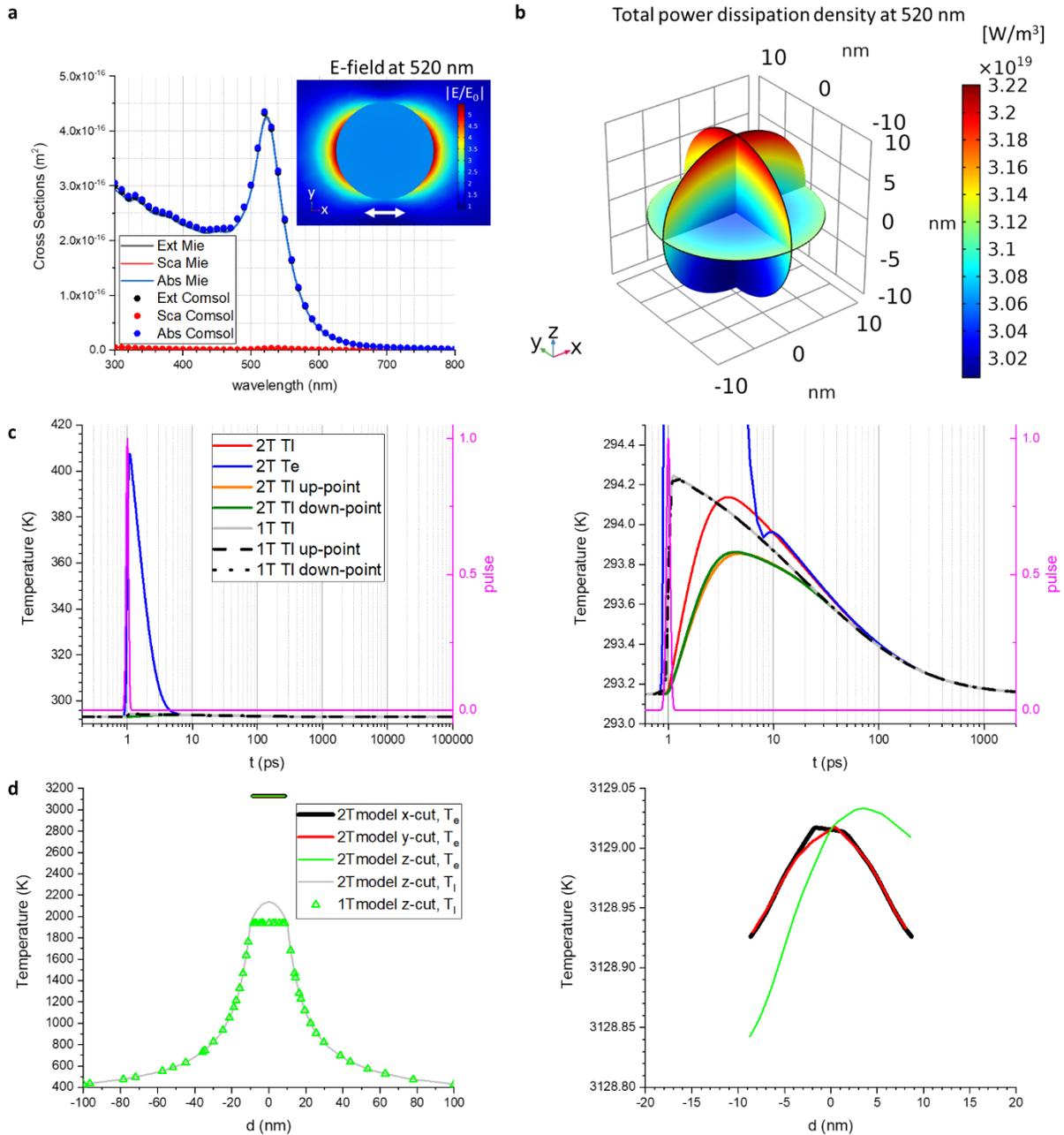

**Figure S3:** (a) Optical cross sections for a 20 nm gold NP in water ($\varepsilon_m = 1.8$), obtained by the finite element difference Comsol simulations used throughout this work and by the benchmark Mie theory. The inset shows the dipolar field enhancement at the plasmon peak 520 nm, with the white arrow indicating the field polarization. $E_0$ is the incident electric field magnitude. (b) 3D distribution of the total power dissipation density when the Au NP is illuminated with k//z and E//x, with a 520 nm excitation wavelength at the plasmon peak of panel (a) and a power of $I_0 =$

$3 \times 10^7$ W/cm². Panels (c) and (d) show the modeling of the heat dissipation comparison of the electronic ($T_e$) and lattice ($T_l$) temperatures solved from the 1T or the 2T models, both (c) temporal and (d) spatial. Both panels (c), full and zoomed-in, show average and superficial temperatures on the left axis, upon photoexcitation (pink right axis) of the total power dissipation shown in (b). Here, the photoexcitation is a gaussian has a duration of 100 fs and is centered at 1 ps. Both panels (d), full and zoomed-in, show the spatial distribution of the temperatures for steady state illumination.

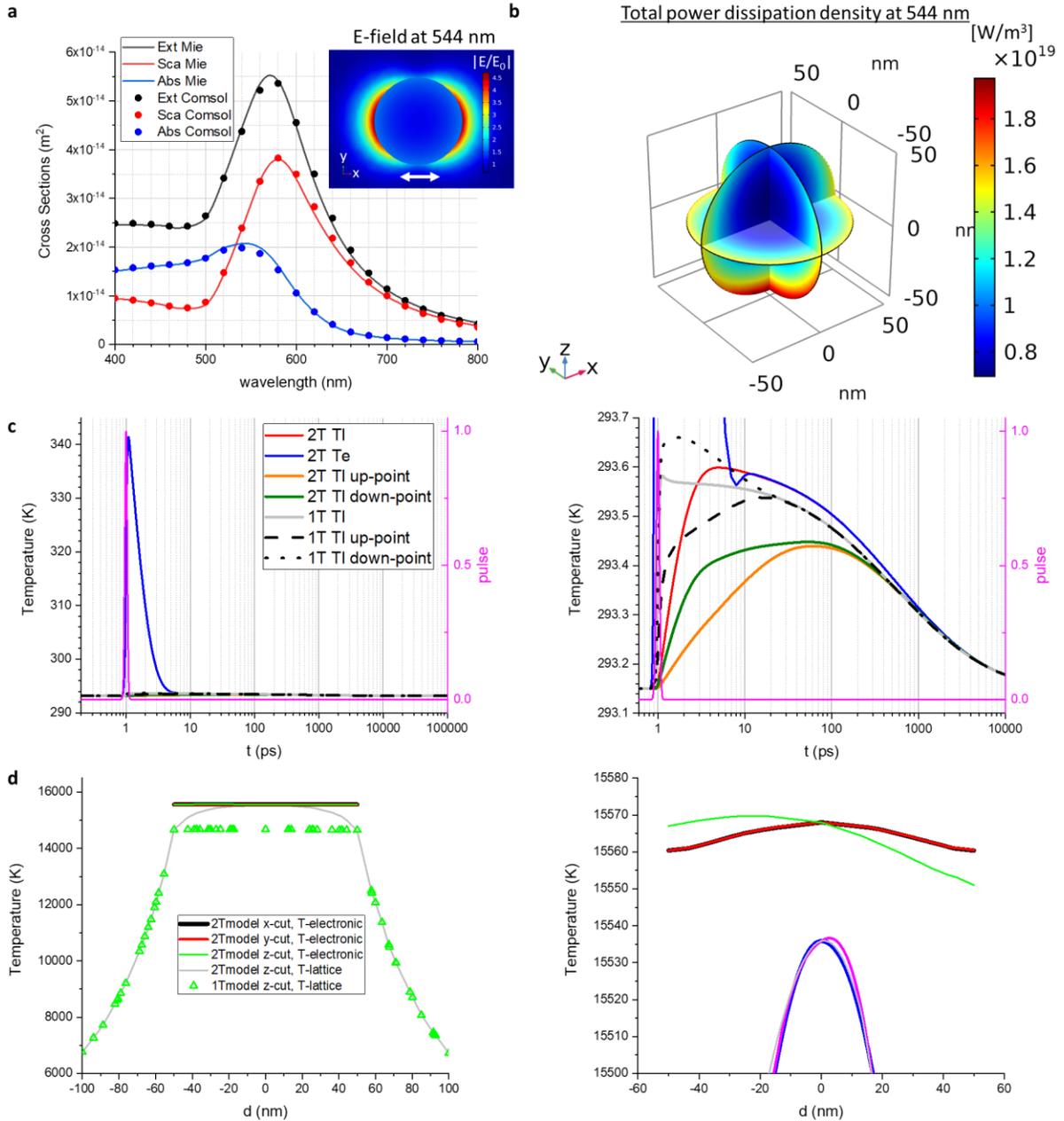

**Figure S4:** (a) Optical cross sections for a 100 nm gold NP in water ($\varepsilon_m = 1.8$), obtained by the finite element difference Comsol simulations used throughout this work and by the benchmark Mie theory. The inset shows the dipolar field enhancement at the plasmon peak 544 nm, with the white arrow indicating the field polarization. $E_0$ is the incident electric field magnitude. (b) 3D distribution of the total power dissipation density when the Au NP is illuminated with k//z and E//x, with a 544 nm excitation wavelength at the plasmon peak of panel (a) and a power of $I_0 = 3 \times 10^7$ W/cm². Panels (c) and (d) show the modeling of the heat dissipation comparison of the electronic ($T_e$) and lattice ($T_l$) temperatures solved from the 1T or the 2T models, both (c) temporal and (d) spatial. Both panels (c), full and zoomed-in, show average and superficial temperatures on the left axis, upon photoexcitation (pink right axis) of the total power dissipation shown in (b). Here, the photoexcitation is a gaussian has a duration of 100 fs and is centered at 1 ps. Both panels (d), full and zoomed-in, show the spatial distribution of the temperatures for steady state illumination.

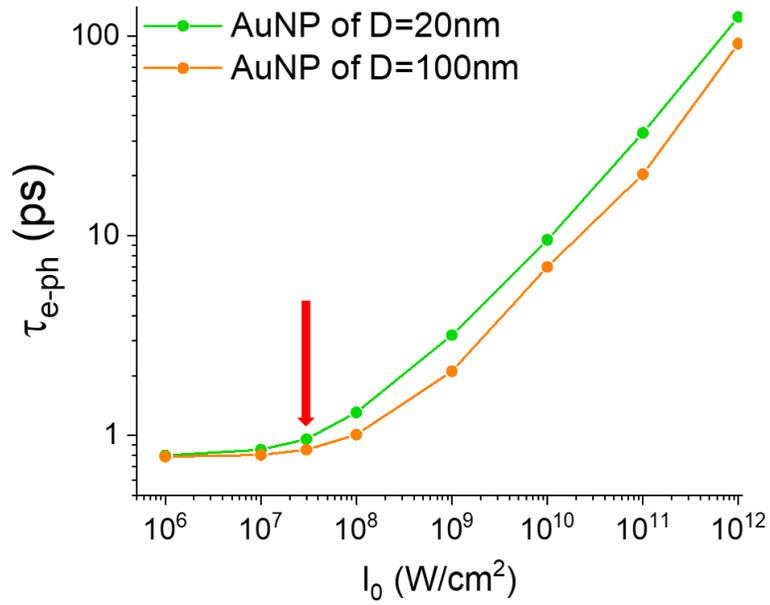

**Figure S5:** Dependence of the electron phonon coupling $\tau_{e-ph}$ on the irradiation intensity $I_0$, for both Au NPs presented in Figs. S3 and S4, both of which were done with $I_0 = 3 \times 10^7$ W/cm$^2$, here indicated with the red arrow.

# III. Metastructure geometries

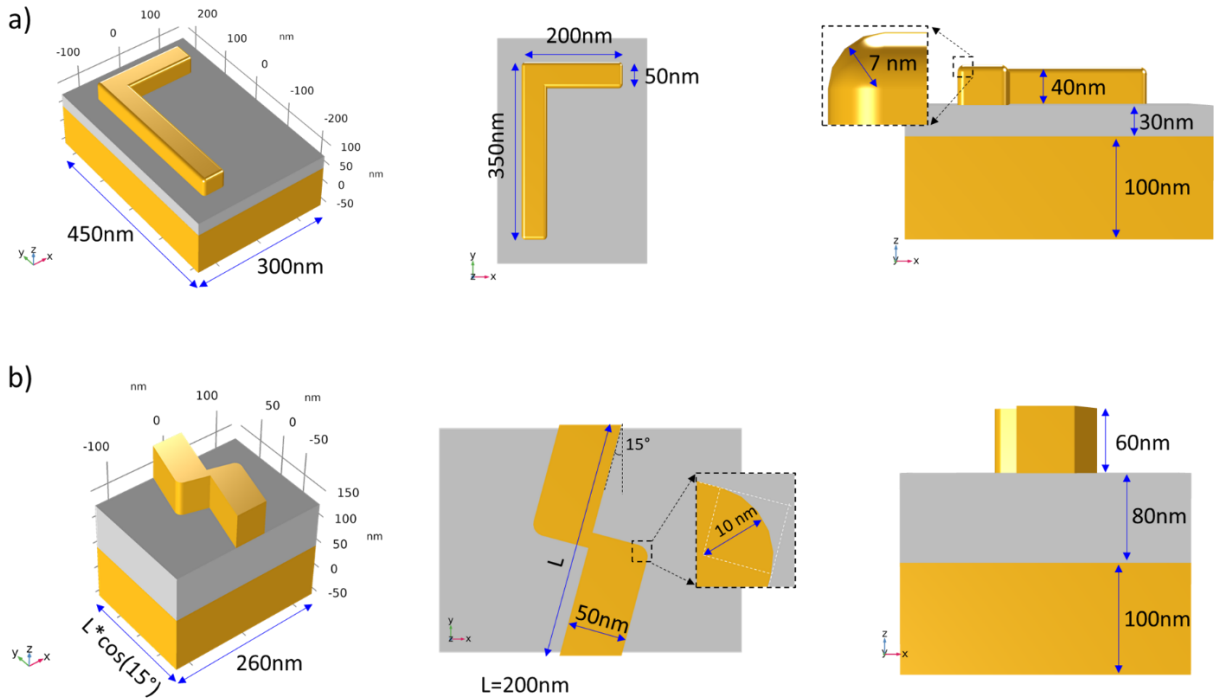

**Figure S6:** Schematics of the chiral structures considered, (a) the chiral L-shape and (b) the chiral wire. Each panel shows different views of the unit cells used to create the metastructures: a panoramic view (left), a top view (middle), and side view (right). Every unit cell is repeated infinitely using periodic boundary conditions throughout the XY plane. The golden regions are gold and the gray regions are glass. Each structure considers infinite perfectly matched layer (PML) of glass towards both the +z and -z directions, not shown here.

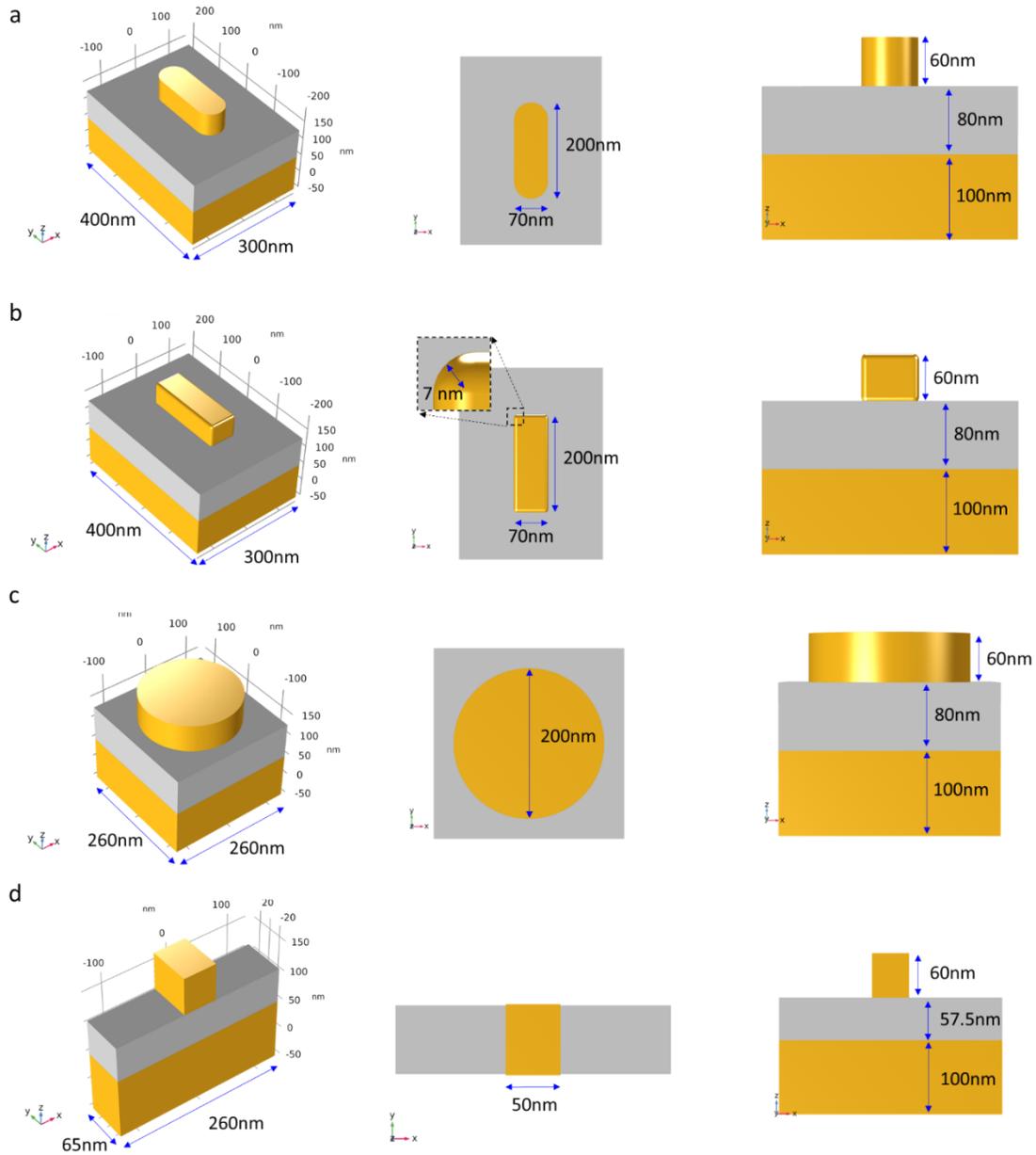

**Figure S7:** Schematics of the achiral structures considered, (a) the rod, (b) the rectangular prism, (c) the disk, and (d) the achiral infinite wire. Each panel shows different views of the unit cells used to create the metastructures: a panoramic view (left), a top view (middle), and side view (right). Every unit cell is repeated infinitely using periodic boundary conditions throughout the XY plane. The golden regions are gold and the gray regions are glass. Each structure considers infinite perfectly matched layer (PML) of glass towards both the +z and -z directions, not shown here.

## IV. Metastructure plasmonic field enhancements

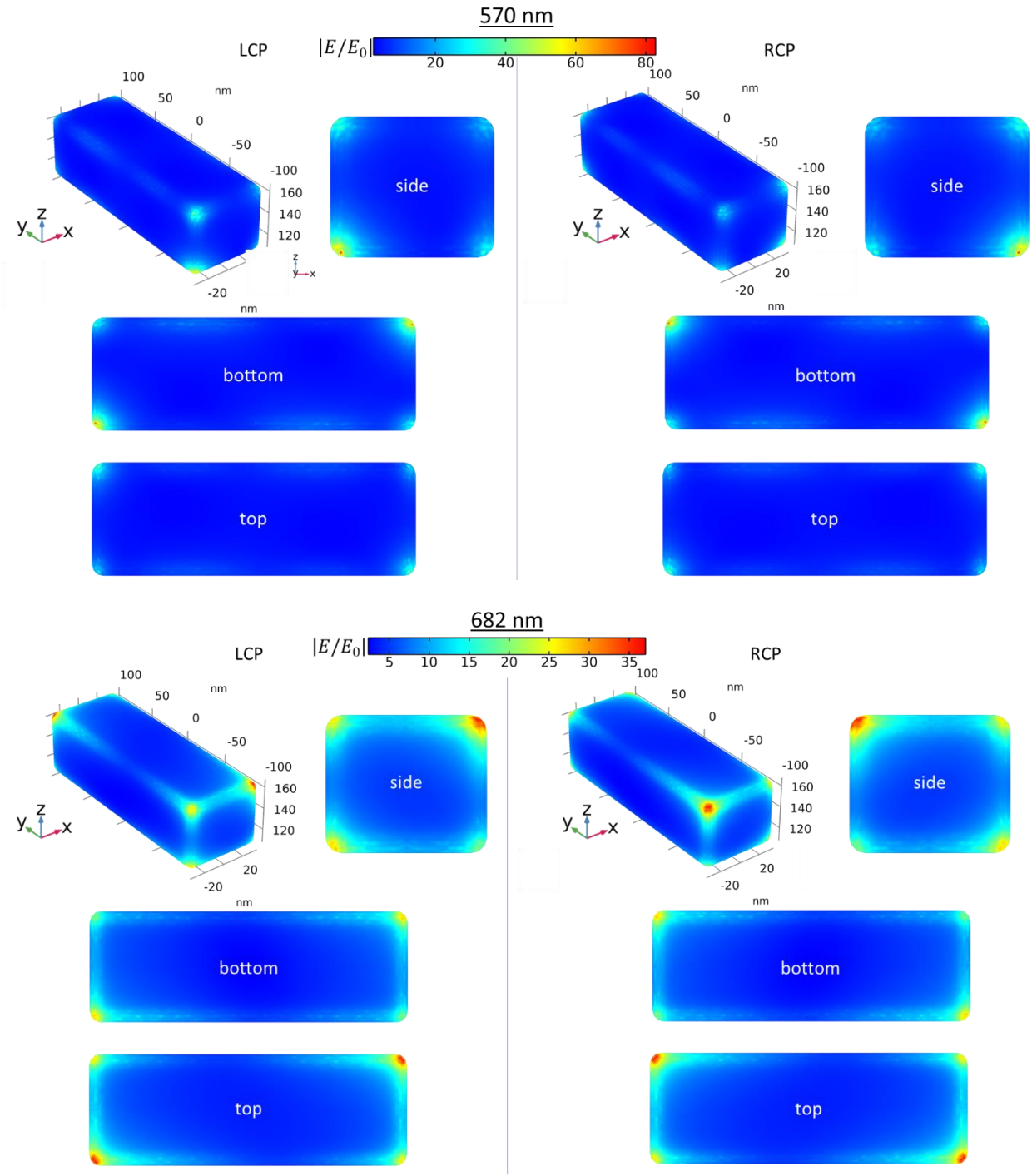

**Figure S8:** Spatial distribution of the electric fields in the gold antenna rectangular prism when the metastructure is excited at both main plasmon resonances, for both LCP and RCP as indicated.

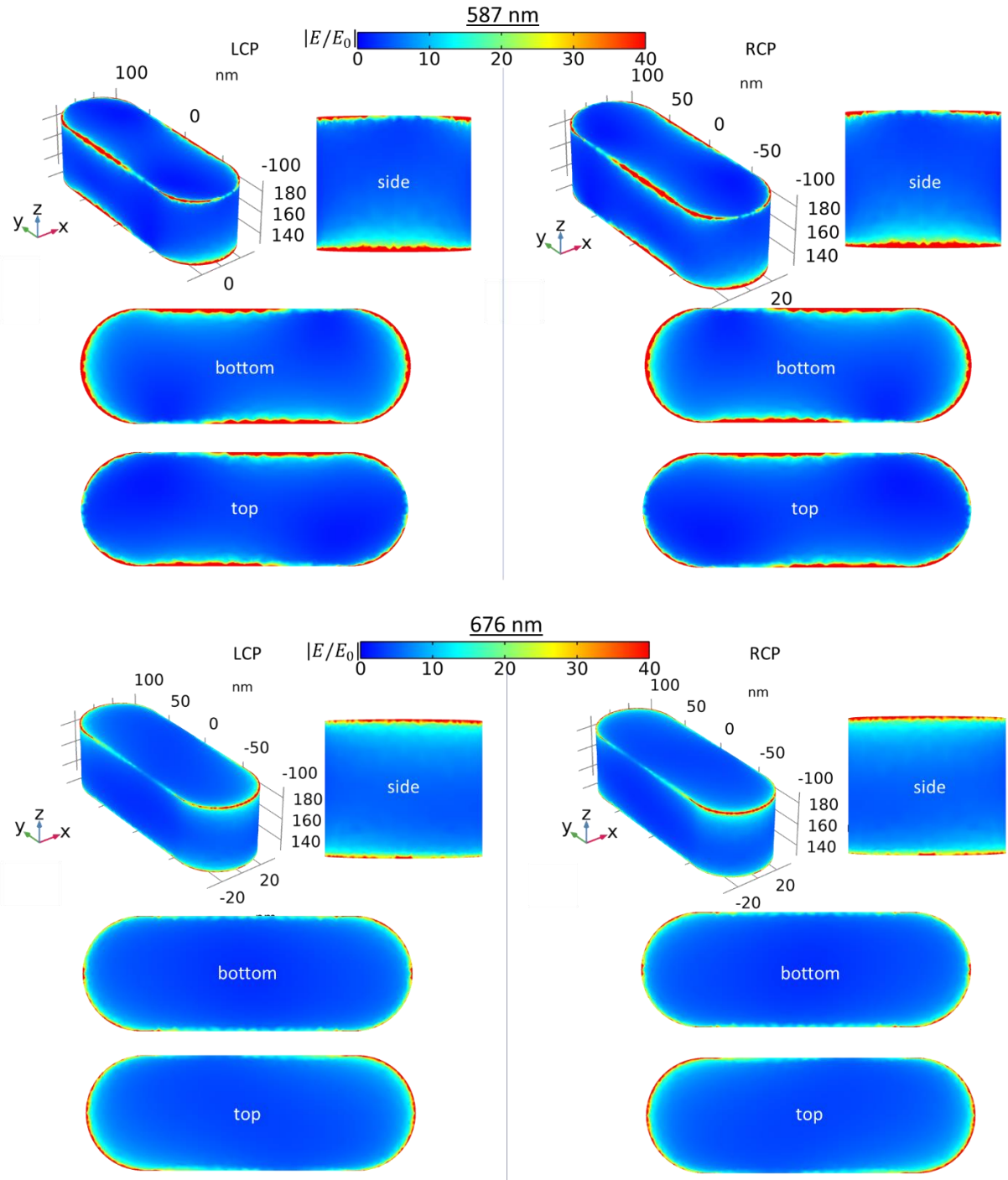

**Figure S9:** Spatial distribution of the electric fields in the gold antenna rod when the metastructure is excited at both main plasmon resonances, for both LCP and RCP as indicated.

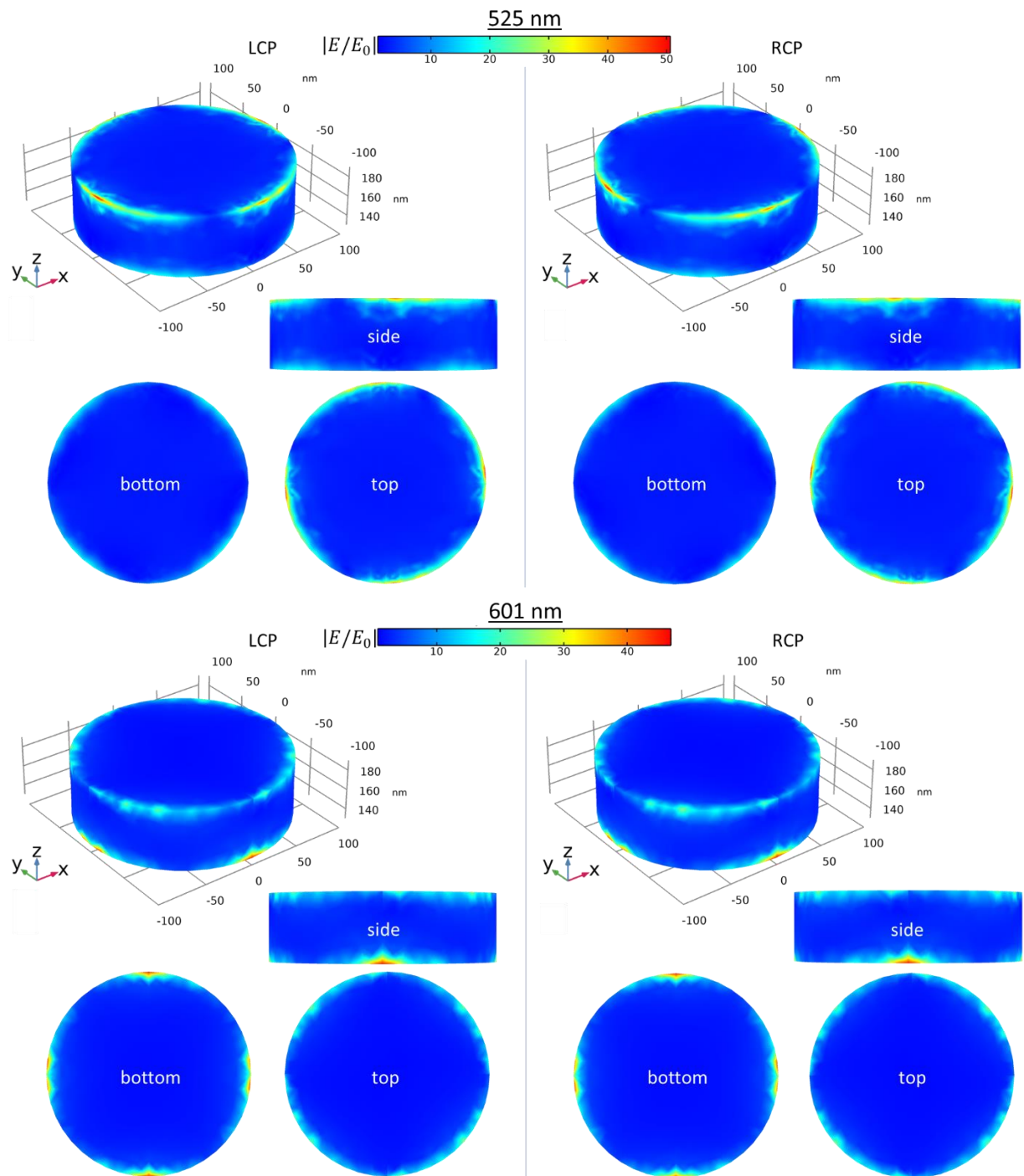

**Figure S10:** Spatial distribution of the electric fields in the gold antenna disk when the metastructure is excited at both main plasmon resonances, for both LCP and RCP as indicated.

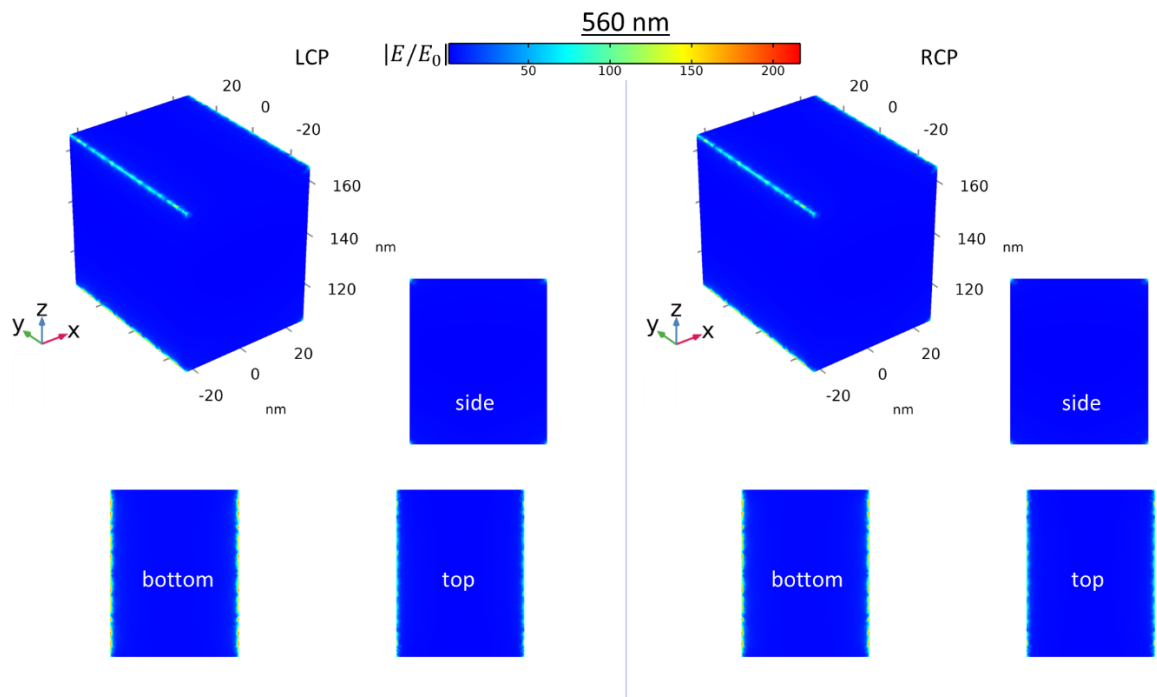

**Figure S11:** Spatial distribution of the electric fields in the achiral wire disk when the metastructure is excited at the main plasmon resonance, for both LCP and RCP as indicated.

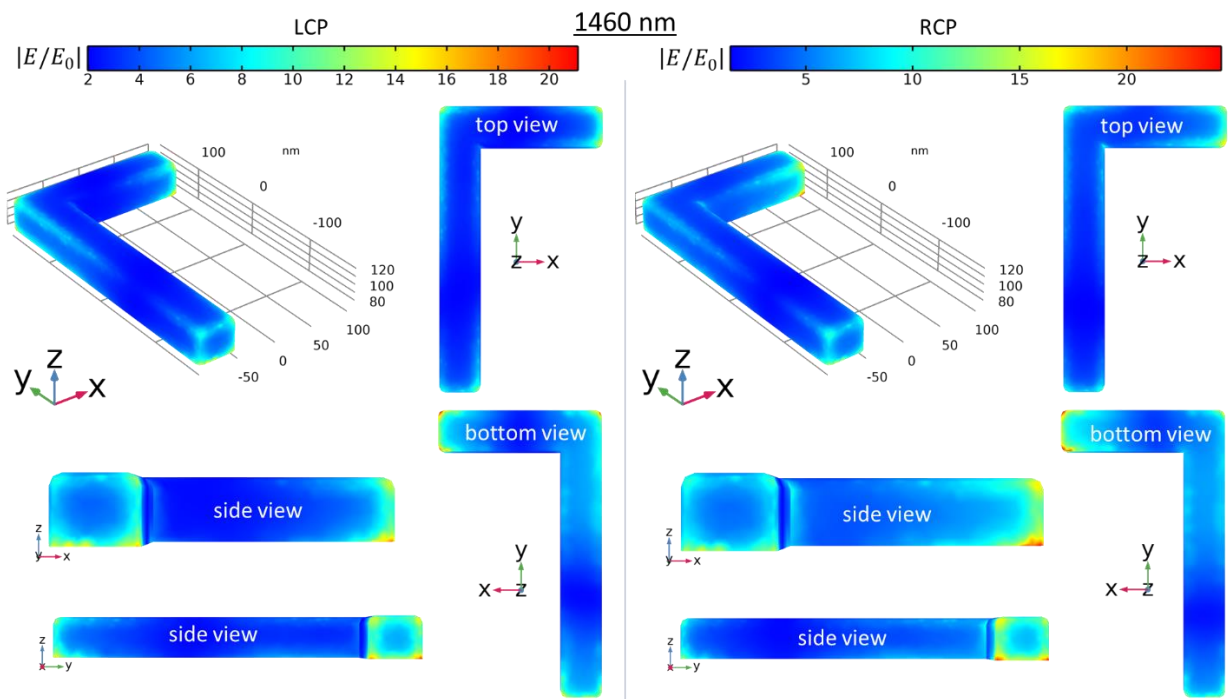

**Figure S12:** Spatial distribution of the electric fields in the gold antenna L-shaped when the metastructure is excited at both main plasmon resonances, for both LCP and RCP as indicated.

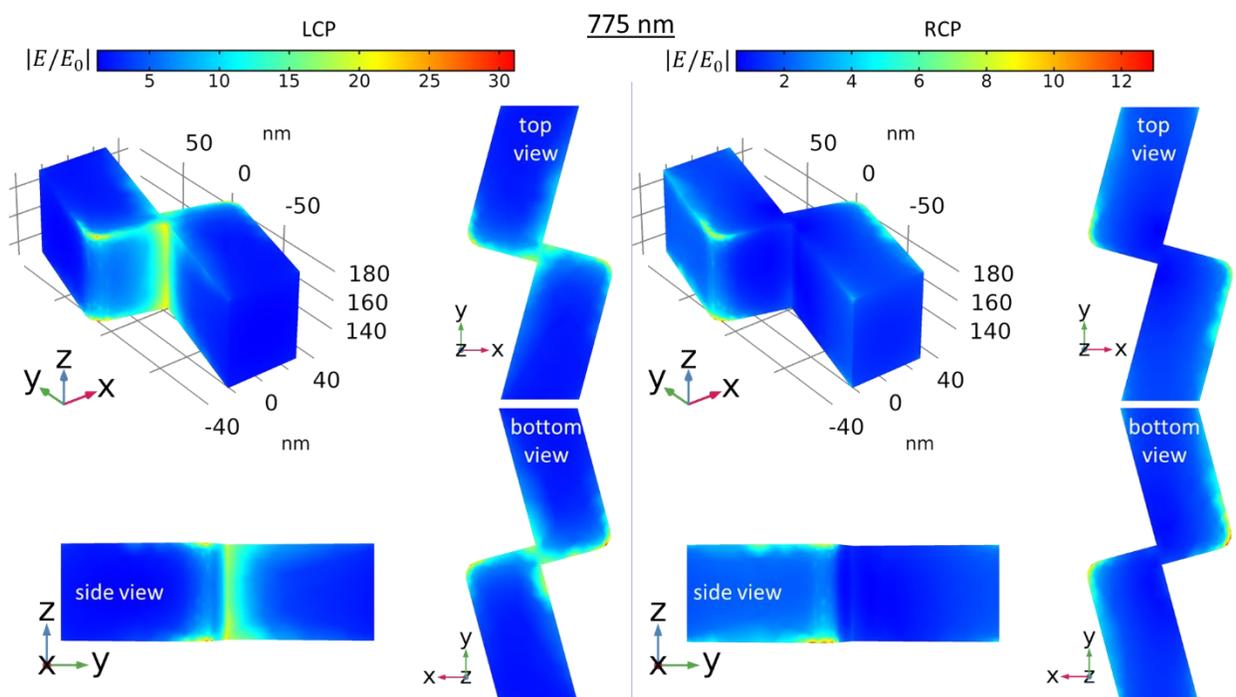

**Figure S13:** Spatial distribution of the electric fields in the gold antenna chiral wire when the metastructure is excited at the main plasmon resonances, for both LCP and RCP as indicated.

# V. Difference between free and mirrored meshes: avoiding false chirality

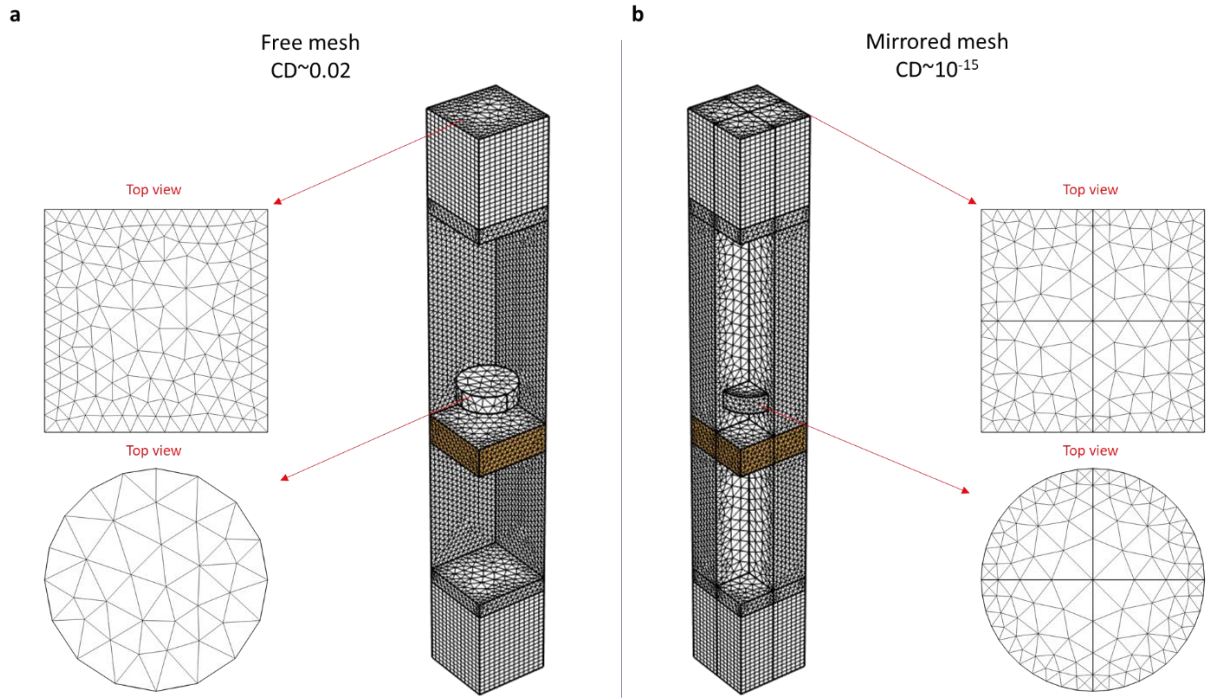

**Figure S14:** Difference between asymmetric and symmetric finite mesh. (a) Asymmetric finite elements create artificial artifact CD ($\sim 10^{-2}$). (b) Perfectly symmetric finite elements do not create artificial artifact CD. The perfect symmetric mesh is created by building "mirrored" meshes. When one reflection plane is used CD$\sim 10^{-8}$ (not shown); whereas when two mirror planes are used CD$\sim 10^{-15}$ (panel b).

# VI. CD and g-factors of the metastructures for CW regime

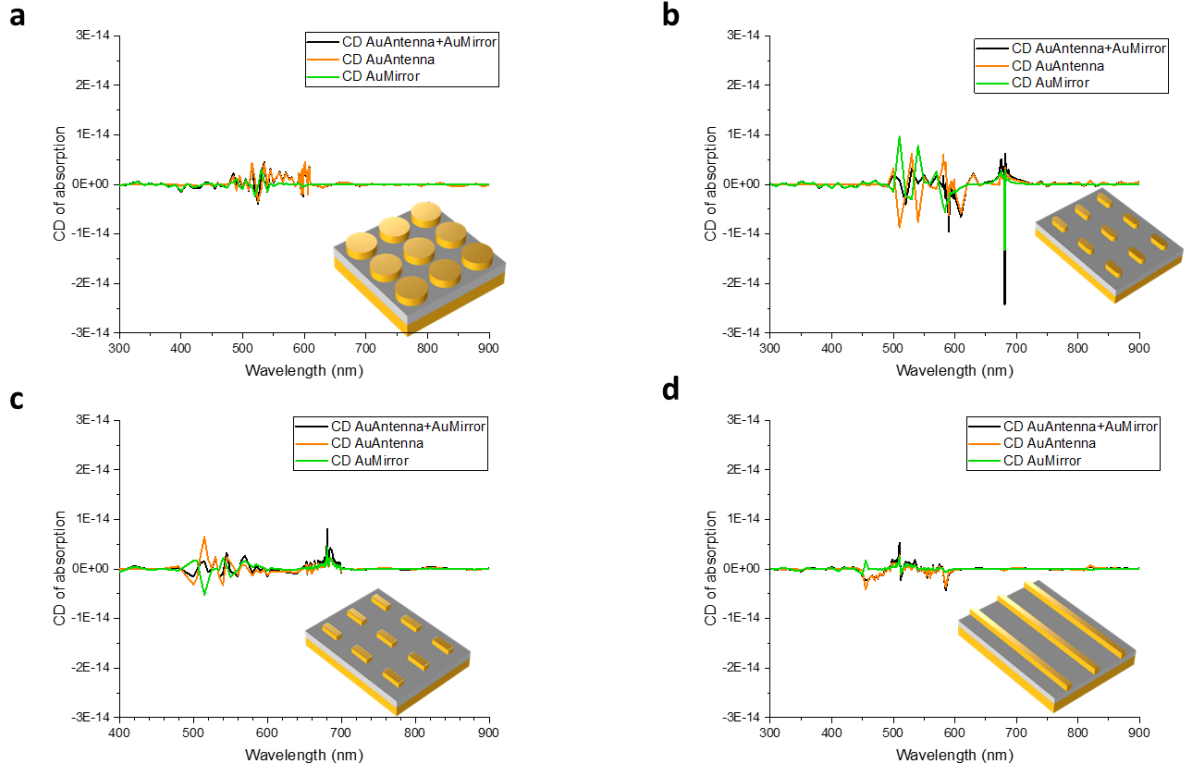

**Figure S15:** Circular dichroism (CD) of the absorption for the achiral structures, all obtained via Eq. 5 from the difference between LCP and RCP absorptions from Fig. 3 of the main text.

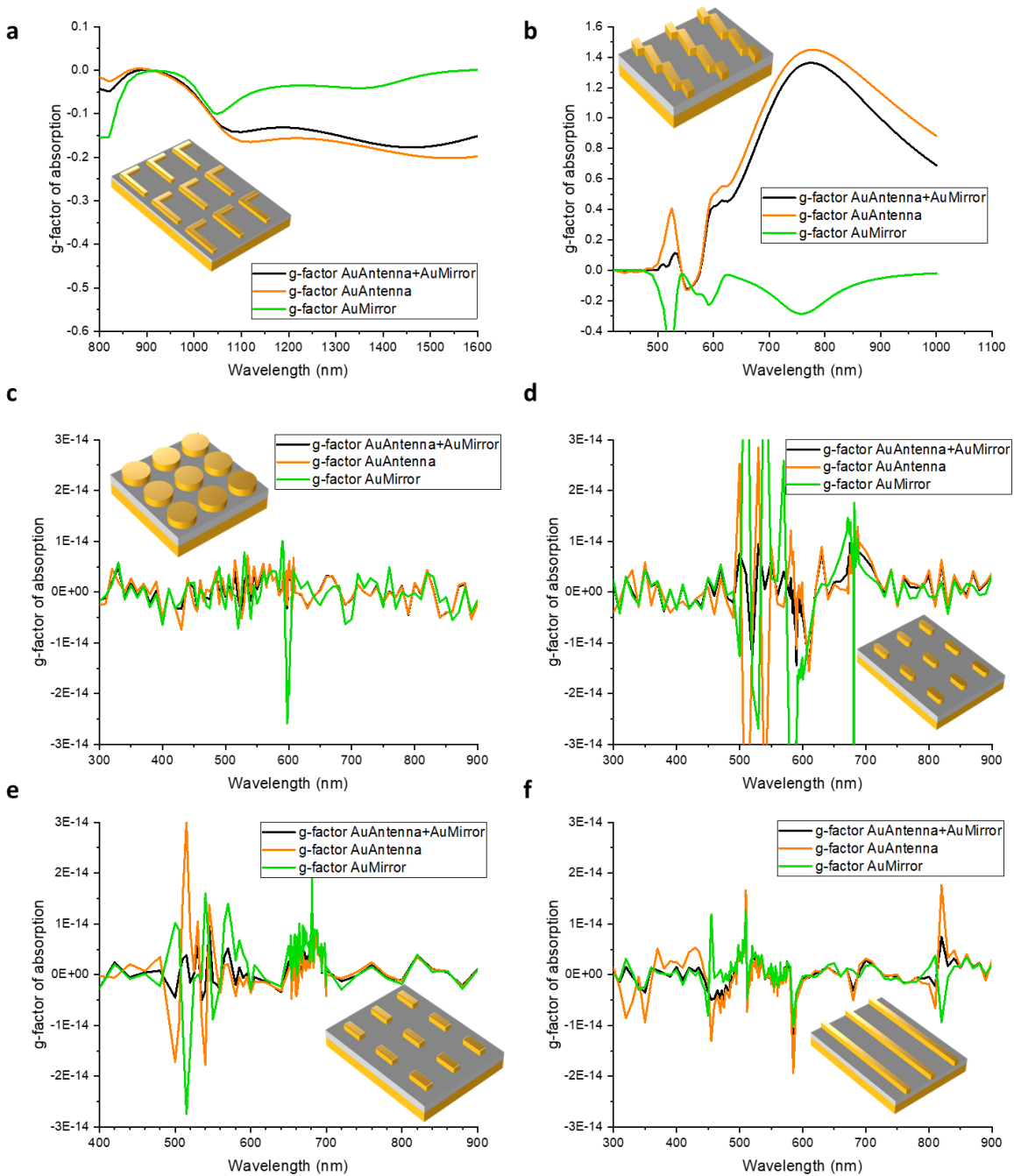

**Figure S16:** g-factor spectra for all chiral and achiral metastructures studied. These plots are obtained via Eq. 6 and the absorptions shown in Figs. 2 and 3 of the main text.

# VII. Pump-probe simulations with chiral metastructures

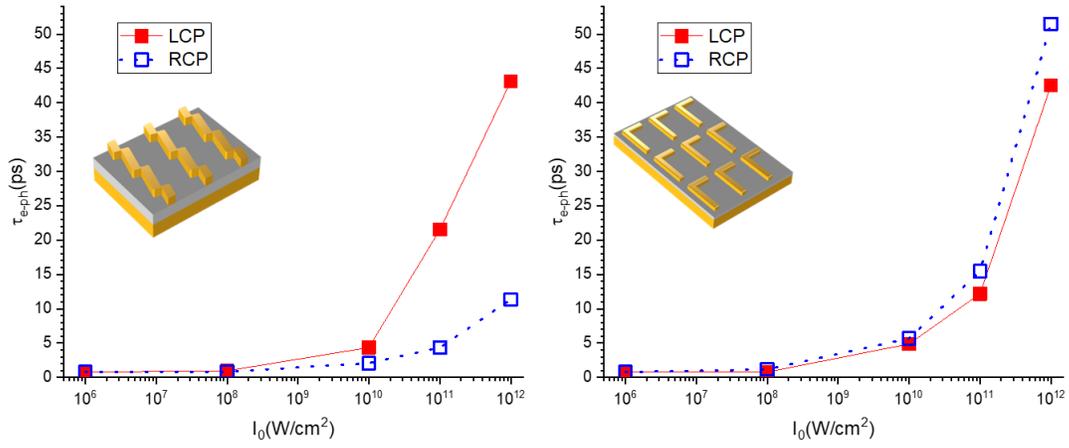

**Figure S17:** Dependence of the electron phonon coupling $\tau_{e-ph}$ on the irradiation intensity $I_0$, for the chiral metastructures.

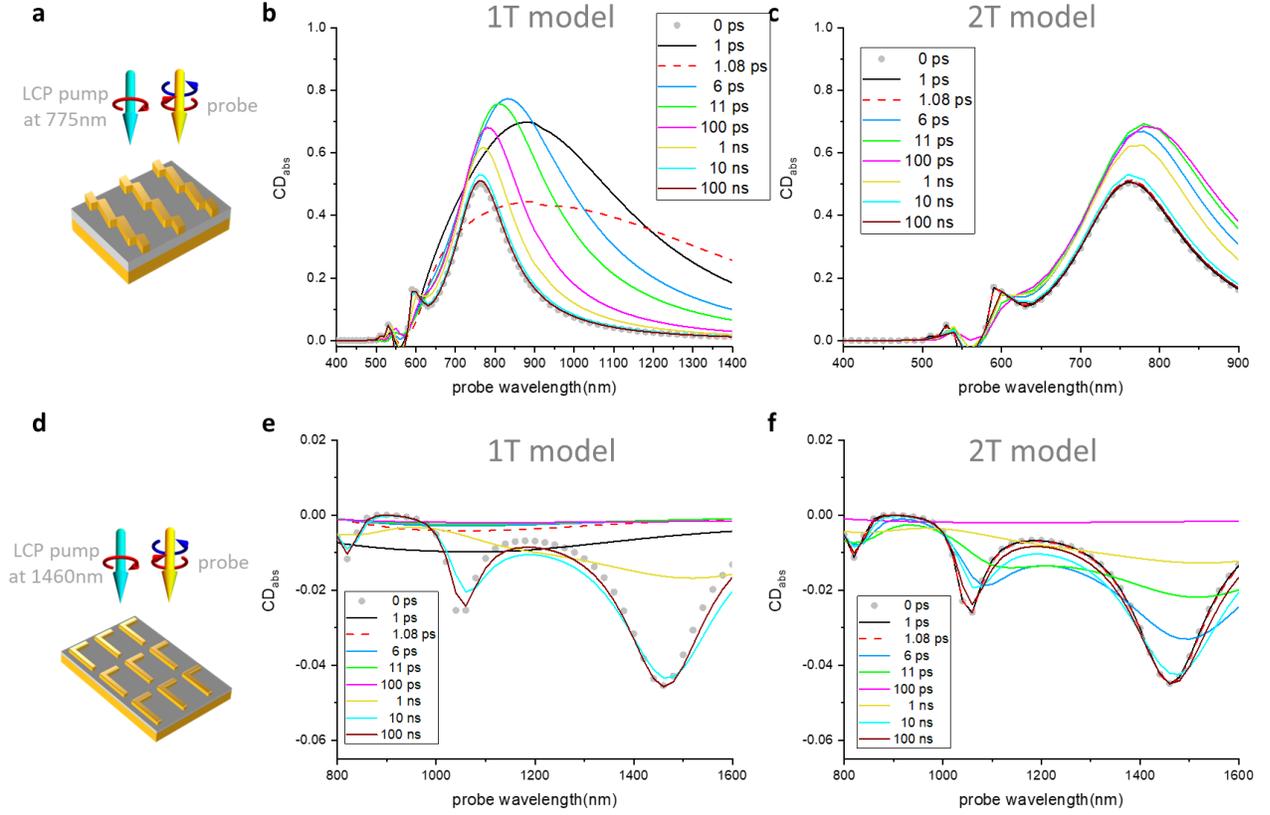

**Figure S18:** CPL photoexcitation and temporal-thermal responses for the chiral wire (a-c) and for the L-shaped metastructures (d-f), both setups for LCP pump and LCP/RCP probes as schematized in (a) and (d), respectively, after a 100 fs pulsed illumination. (b) and (e) are individual $CD(t)$ curves for the times there specified, obtained via the 1T model. (c) and (f) are the same description as (b) and (e), but now for the 2T model. For all panels we also include the CW CD shown in Figs. 2b and 2e (solid gray circles). The differences between the CD at any specific time $t$ (color lines) and the CD at with CW (solid gray circles), defined as $\Delta CD_{abs} = CD(t) - CD_0$, are shown in Fig. 4 from the main text. The chiral wire was simulated with $I_0 = 10^{11}$ W/cm², and the L-shaped with $I_0 = 10^{11}$ W/cm², both with a CPL beam $\vec{E} // -\vec{k}_z$.

For the achiral metastructures, we see that when observing results for the 1T model, both metastructures show their maximum $\Delta CD_{abs}(t)$ exactly at ($t_0 = 1$ ps, black line in Figs. S19c and S19g) and a few fs after ($t = 1.08$ ps, dashed red line in Figs. S19c and S19g) irradiation, and then for each increasing time, $\Delta CD_{abs}(t)$ decreases monotonously until recovering the CW response after several ns, in our cases the largest calculated time ($t = 100$ ns, wine color curves in Figs. S19c and S19g).

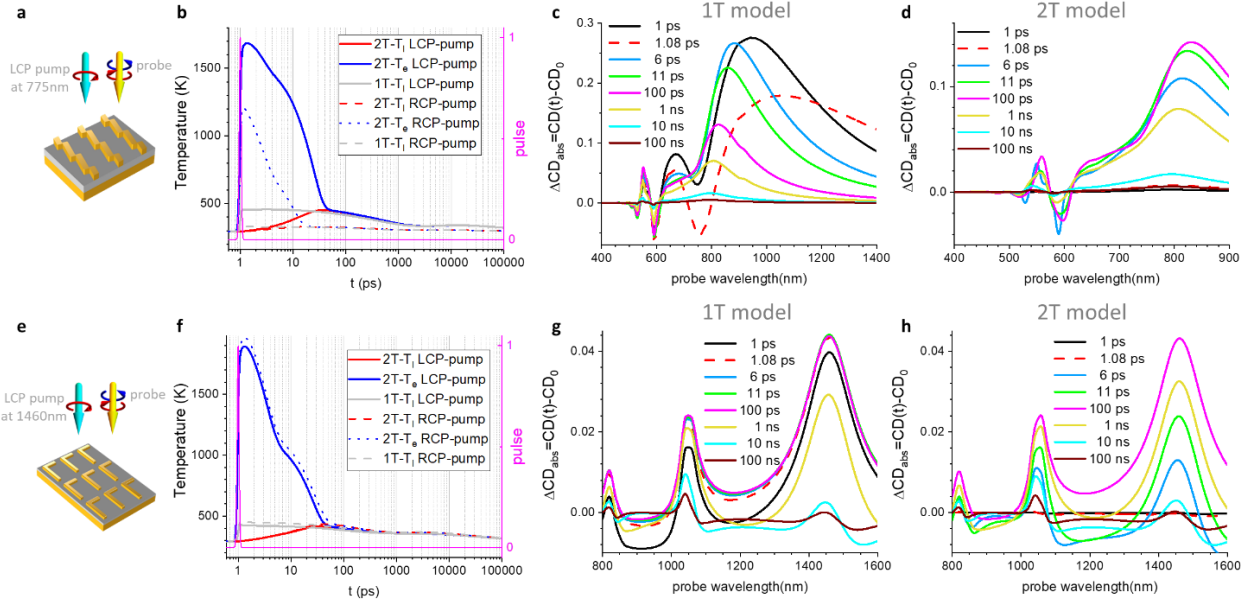

**Figure S19:** CPL photoexcitation and temporal-thermal responses for the chiral wire (a-d) and for the L-shaped metastructures (e-h), both setups for LCP pump and LCP/RCP probes as schematized in (a) and (e), respectively. (b) and (f) are the volumetric temperature averages over the entire structure for both 1T (gray) and 2T (red for $T_l$, blue for $T_e$) models, for either pumping LCP (solid lines) or RCP (dashed lines), after the 100 fs pulsed illumination (pink line, right axis). (c) and (g) for the 1T model, and (d) and (h) for the 2T model, are the CD difference $\Delta CD_{abs} = CD(t) - CD_0$, this is between the CD at any specific time $t$ and the CD at with CW. Individual $CD(t)$ curves are provided in Fig. S18. Both were simulated with $I_0 = 10^{11}$ W/cm$^2$, and pumped at the plasmon frequency described in the CW regime (see Fig. 2) with a CPL beam $\vec{E} // -k_z$.

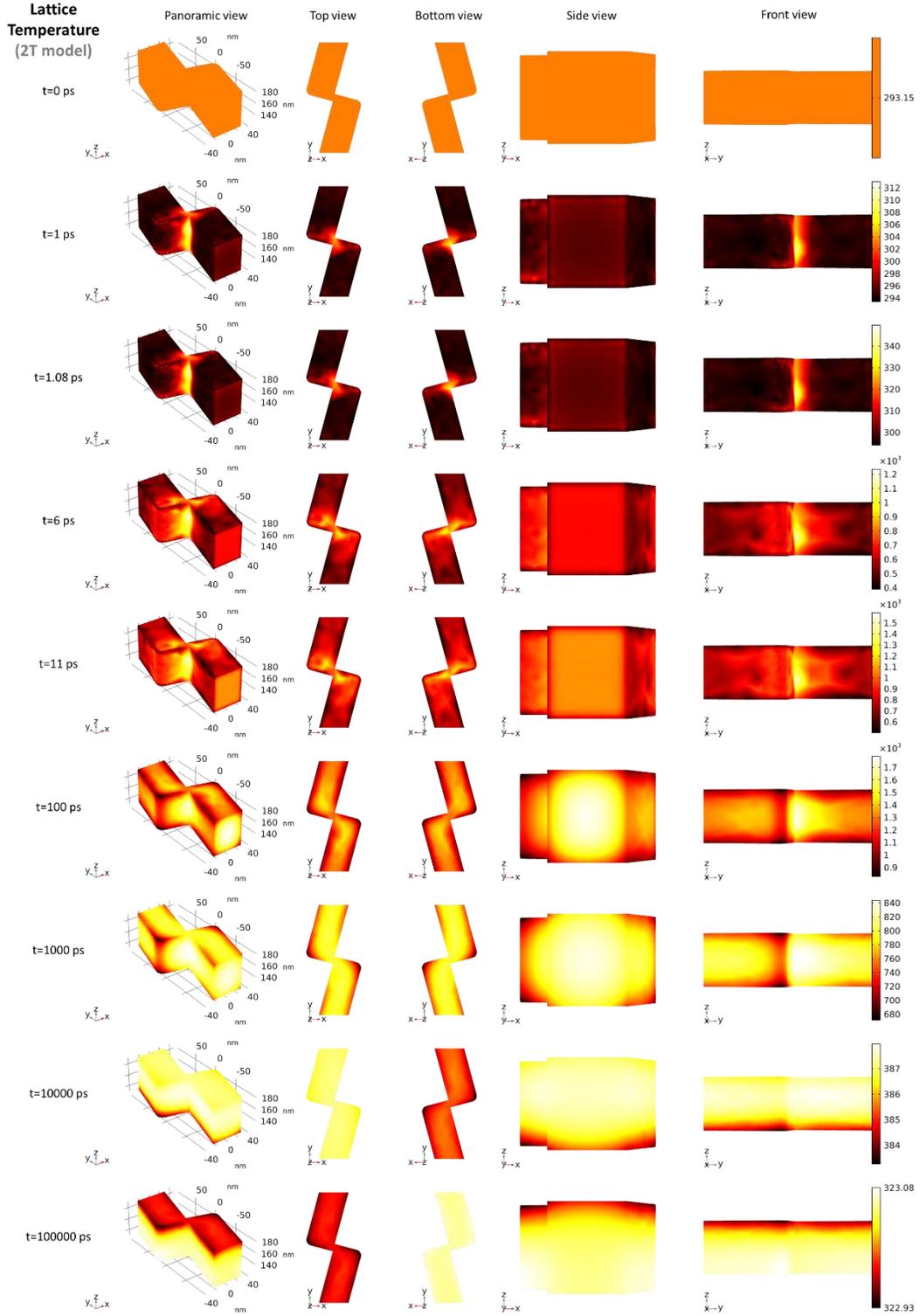

**Figure S20:** Lattice temperature heat maps for the chiral wire antenna in the metastructure configuration, for the solutions of the 2T model with LCP pump. The metastructure was simulated with $I_0 = 10^{11}$ W/cm$^2$, with an LCP beam $\vec{E} \mathbin{/\mkern-6mu/} -k_z$. All bar units are in K.

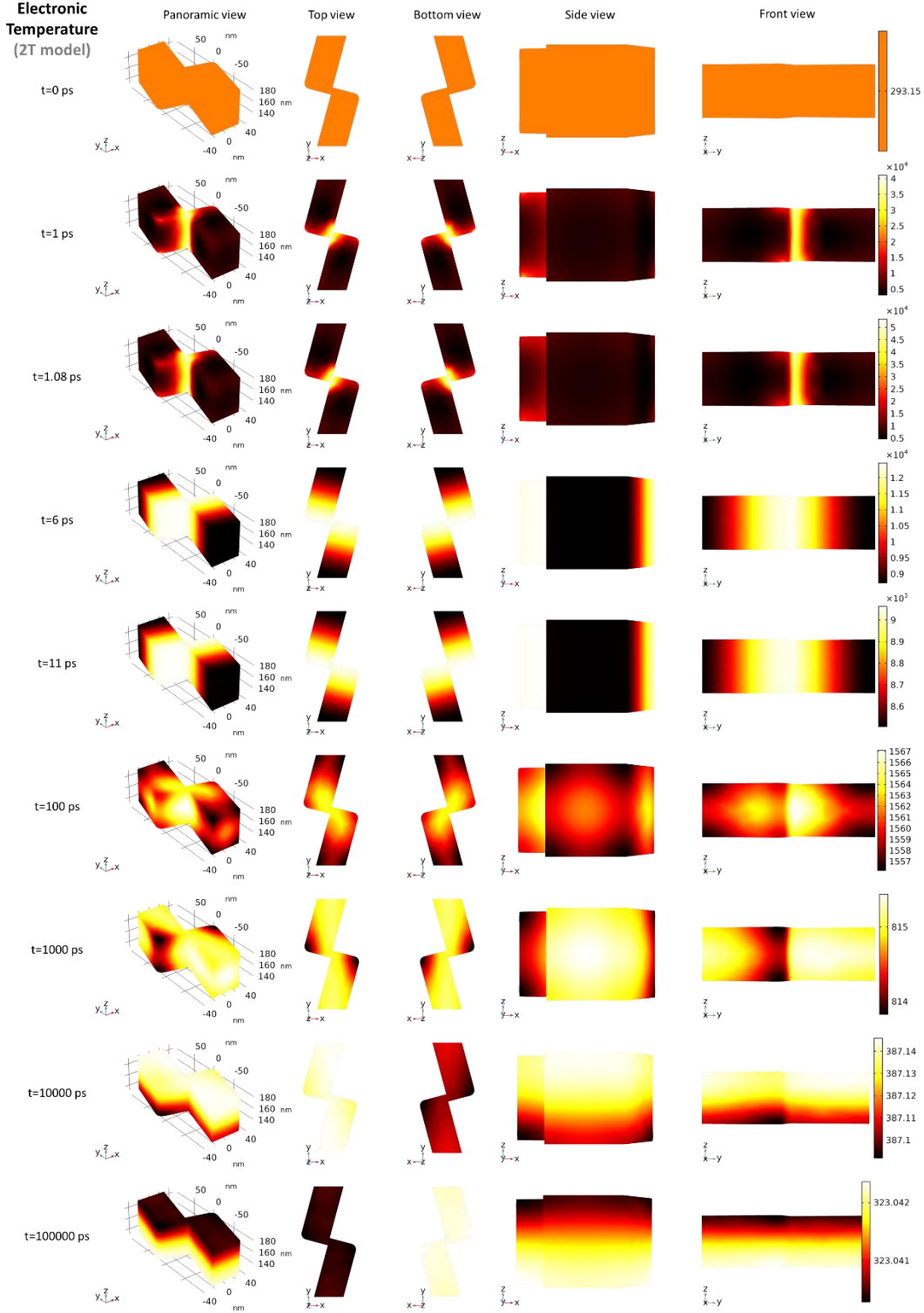

**Figure** S21: Electronic temperature heat maps for the chiral wire antenna in the metastructure configuration, for the solutions of the 2T model with LCP pump. The metastructure was simulated with $I_0 = 10^{11}$ W/cm$^2$, with an LCP beam $\vec{E} // -k_z$. All bar units are in K.

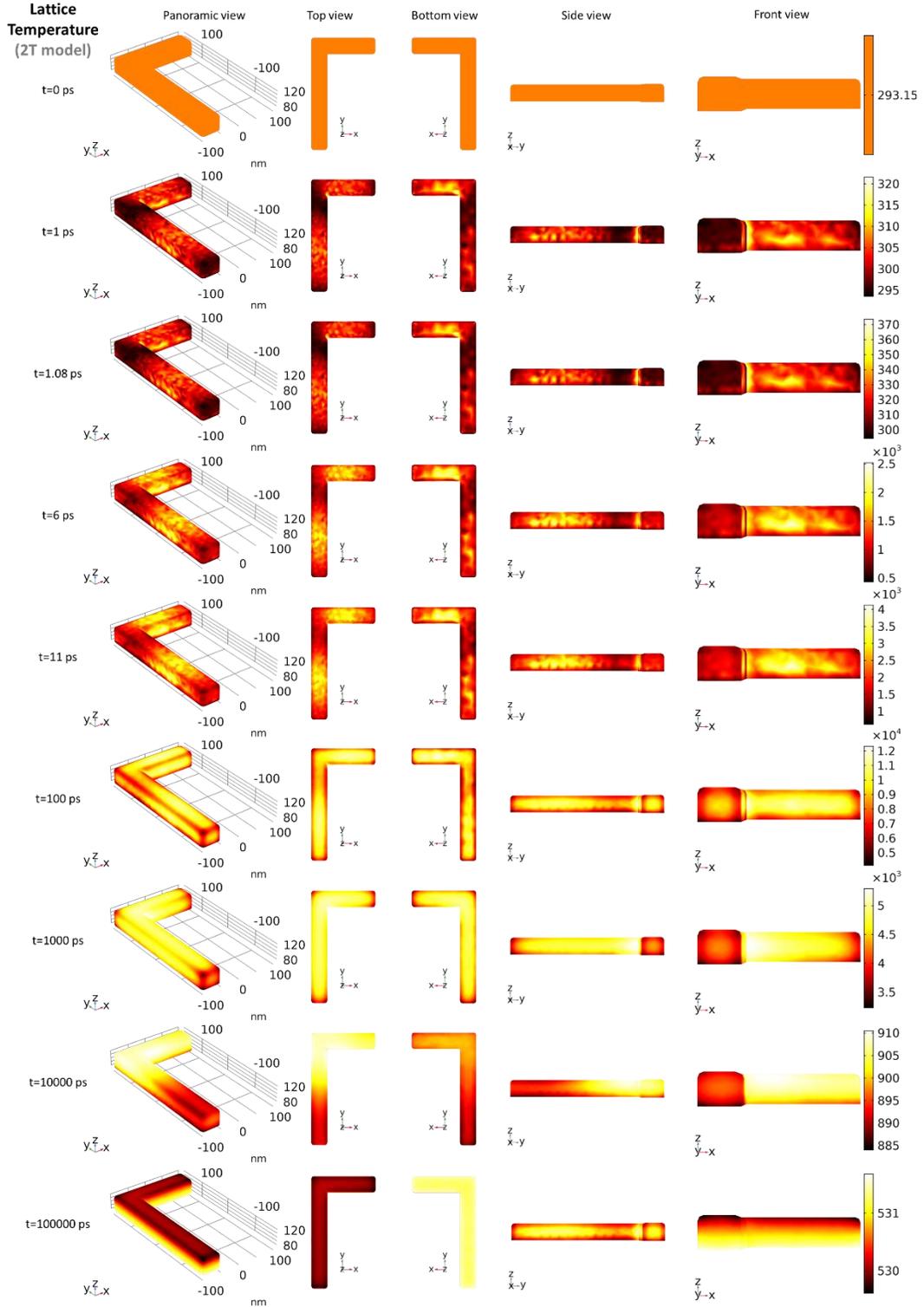

**Figure S22:** Lattice temperature heat maps for the chiral L-shaped antenna in the metastructure configuration, for the solutions of the 2T model with LCP pump. The metastructure was simulated with $I_0 = 10^{12}$ W/cm$^2$, with an LCP beam $\vec{E} // -k_z$. All bar units are in K.

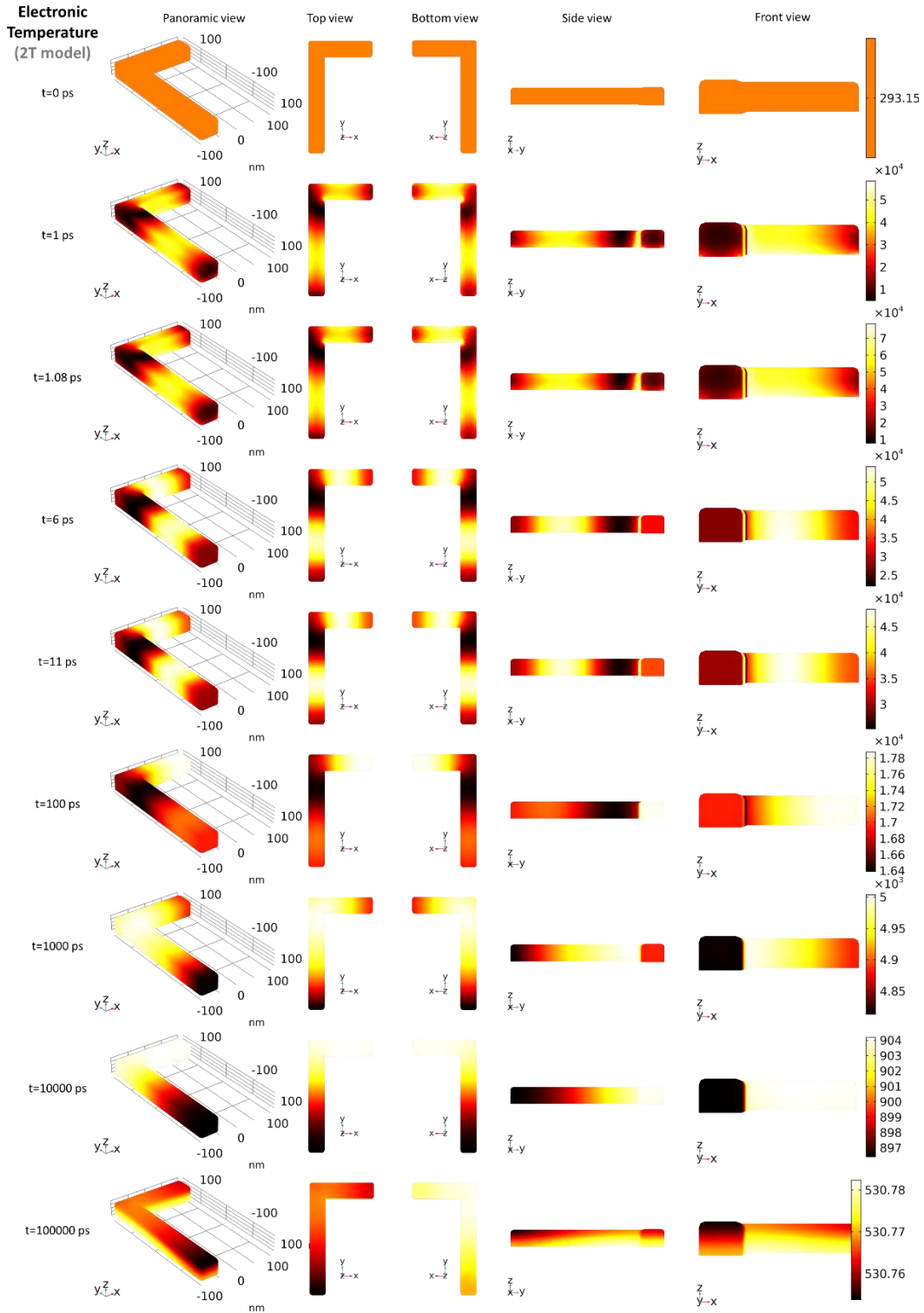

**Figure S23:** Electronic temperature heat maps for the chiral L-shaped antenna in the metastructure configuration, for the solutions of the 2T model with LCP pump. The metastructure was simulated with $I_0 = 10^{12}$ W/cm², with an LCP beam $\vec{E} // -k_z$. All bar units are in K.

## VIII. Pump-probe simulations with achiral metastructures

When analyzing the 1T-model (Figs. S24 c,f,i,l) and similarly as with the chiral structures (Figs. S18b and S18e), the probe CD shows only some electromagnetic "artifact" CD at ($t_0 = 1$ ps, black lines) and a few fs after ($t = 1.08$ ps, dashed red lines) irradiation, due to the chirality of the electrons excited in the structure. Note that for these times, whether the disk and the achiral wire show a $|CD(t)| \approx 10^{-4} - 10^{-5}$, the rod and rectangular prism show significantly larger CD, with even some non-negligible CD for 5 and 100 ps, highlighting that the 1T model is capable of catching the chirality imprinting at early time scales.

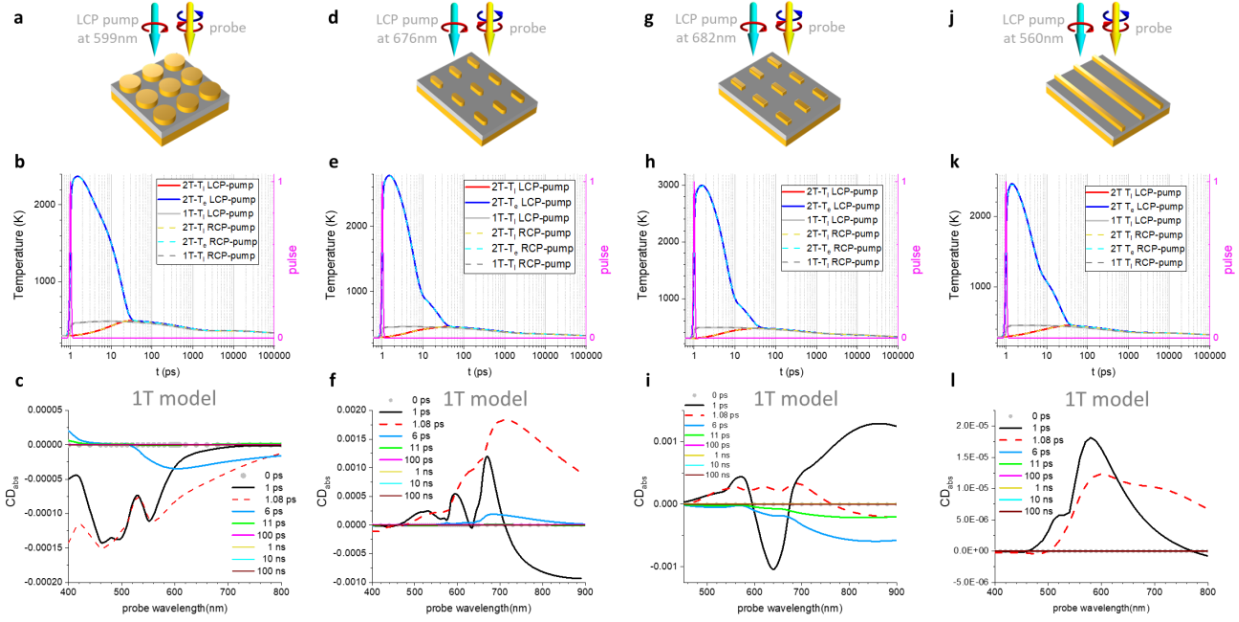

**Figure S24:** 1T-dynamics of the CPL photoexcitation and temporal-thermal responses for the disks (a-c), for the rods (d-f), for the rectangular prisms (g-i), and for the achiral wire (j-l) metastructures, as schematized in top panels, respectively, all for LCP pump and LCP/RCP probes. (b,e,h,k) are the volumetric temperature averages over the entire structure for both 1T (grays) and 2T (red/yellow for $T_l$, blue/cyan for $T_e$) models, for either pumping LCP (solid lines) or RCP (dashed lines), after the 100 fs pulsed illumination (pink line, right axis). (c,f,i,l) are individual $CD(t)$ curves for the times there specified, obtained via the 1T model (See Fig. 5 for the description with the 2T model). For all panels we also include the CW CD, which is inherently zero, shown in Fig. S15. All structures were simulated with $I_0 = 10^{11}$ W/cm², and pumped at the plasmon frequency described in the CW regime (see Fig. 3) with a CPL beam $\vec{E} \parallel -\vec{k}_z$.

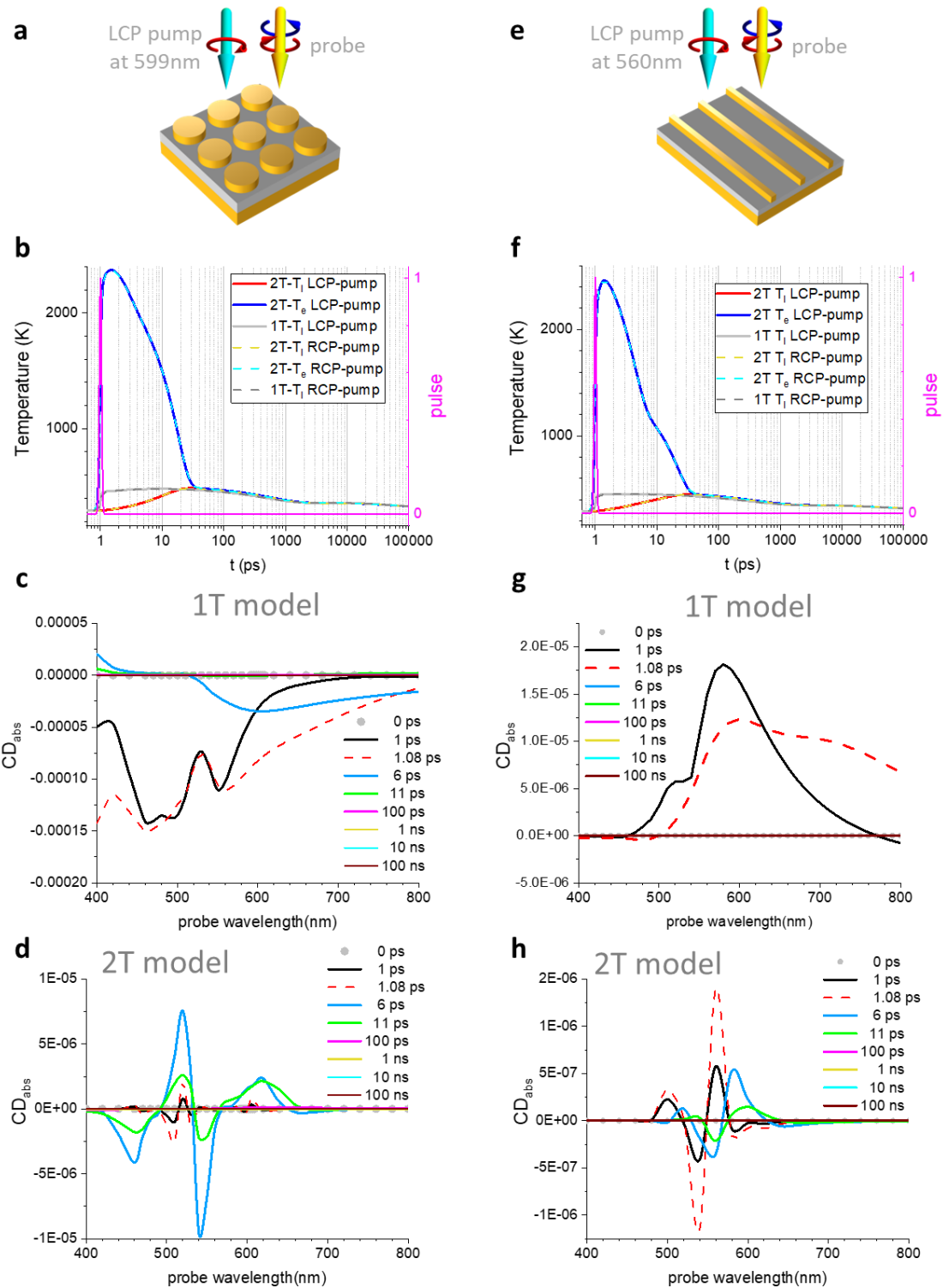

**Figure S25:** 1T vs 2T dynamics of the CPL photoexcitation and temporal-thermal responses for the disks (a-d) and for the achiral wire (e-h) metastructures, as schematized in top panels, respectively, all for LCP pump and LCP/RCP probes. (b,f) are the volumetric temperature

averages over the entire structure for both 1T (grays) and 2T (red/yellow for $T_l$, blue/cyan for $T_e$) models, for either pumping LCP (solid lines) or RCP (dashed lines), after the 100 fs pulsed illumination (pink line, right axis). (c,g) are individual $CD(t)$ curves for the times there specified, obtained via the 1T model; (d,h) are the same but for the 2T model. For all panels we also include the CW CD, which is inherently zero, shown in Fig. S15. All structures were simulated with $I_0 = 10^{11}$ W/cm², and pumped at the plasmon frequency described in the CW regime (see Fig. 3) with a CPL beam $\vec{E}//-\vec{k}_z$.

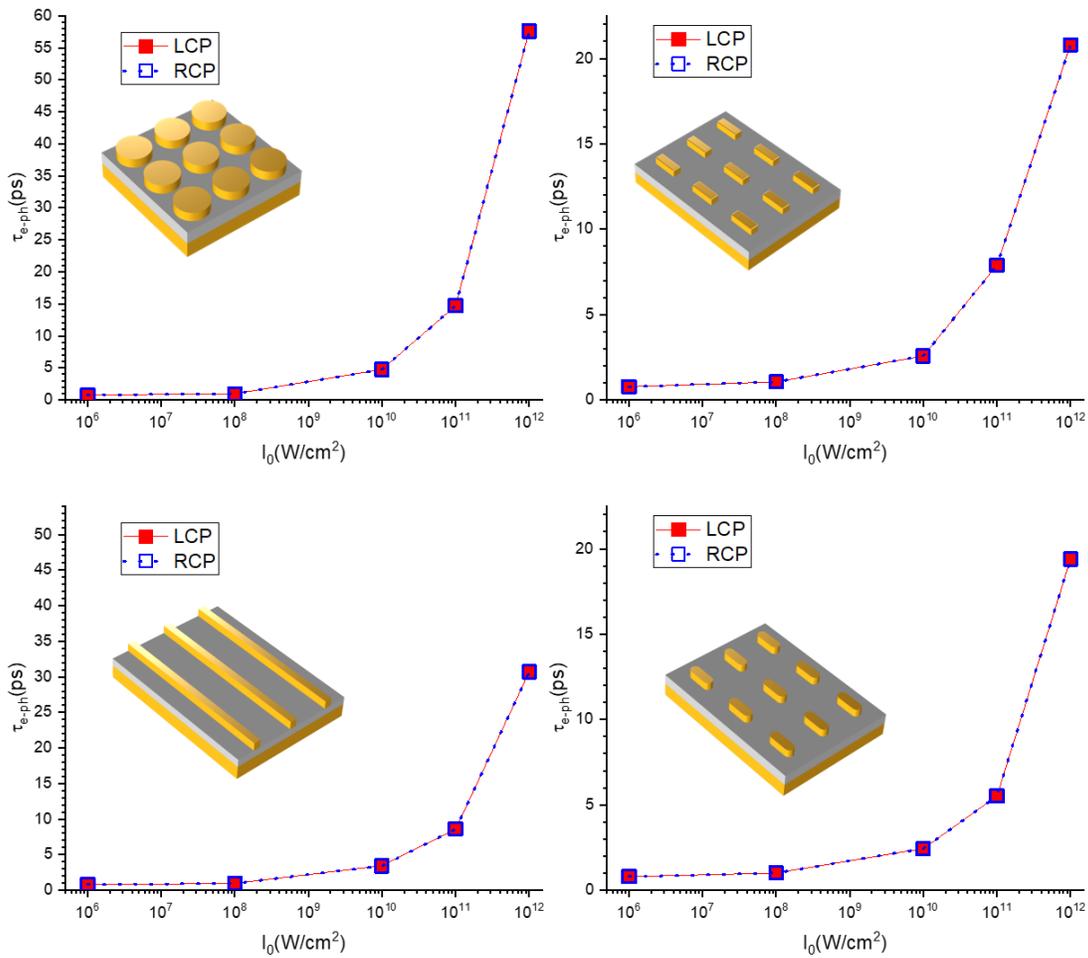

**Figure S26:** Dependence of the electron phonon coupling $\tau_{e-ph}$ on the irradiation intensity $I_0$, for the achiral metastructures.

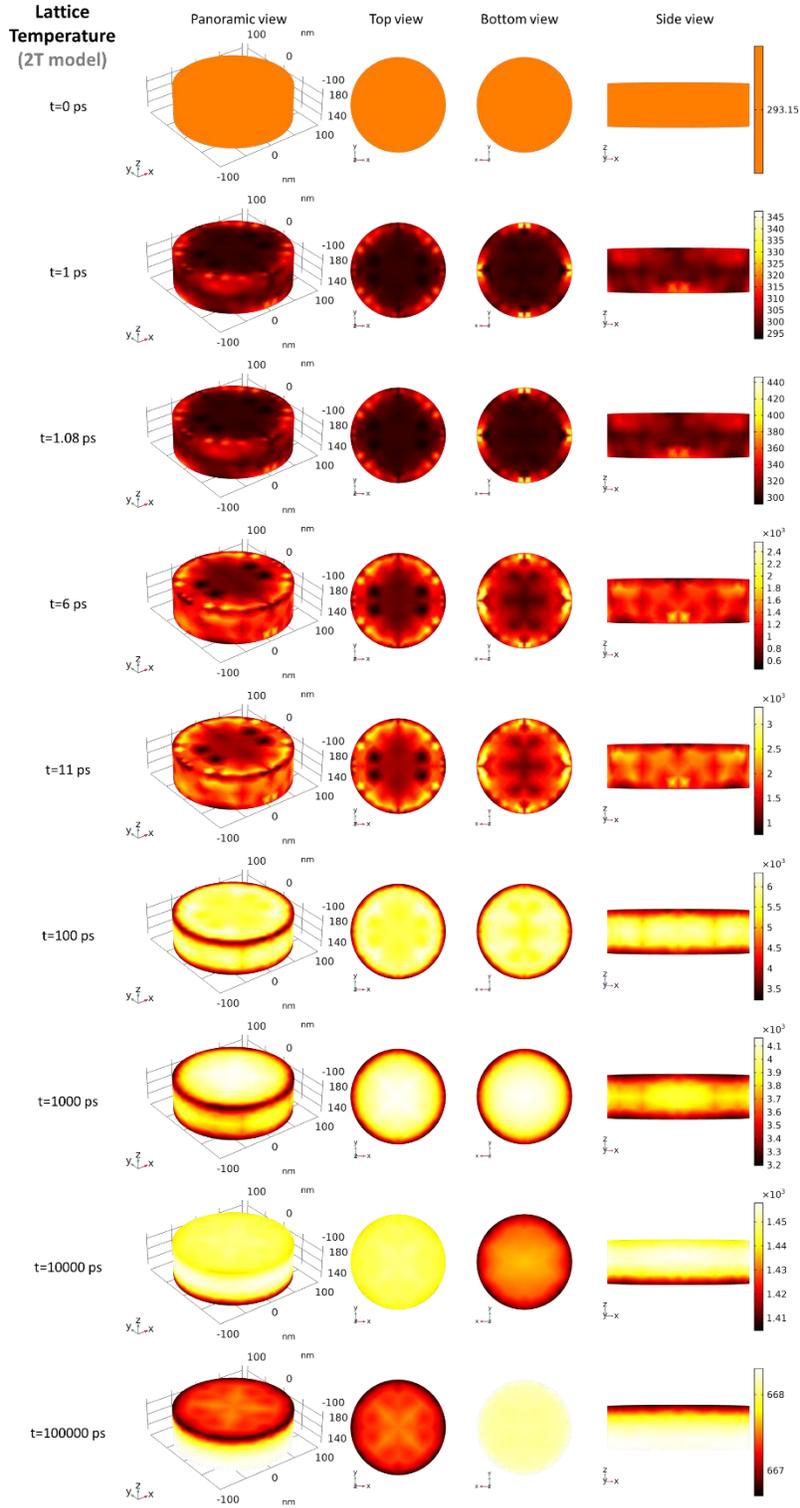

**Figure S27:** Lattice temperature heat maps for the disk antenna in the metastructure configuration, for the solutions of the 2T model with LCP pump. The metastructure was simulated with $I_0 = 10^{12}$ W/cm$^2$, with an LCP beam $\vec{E} // -k_z$. All bar units are in K.

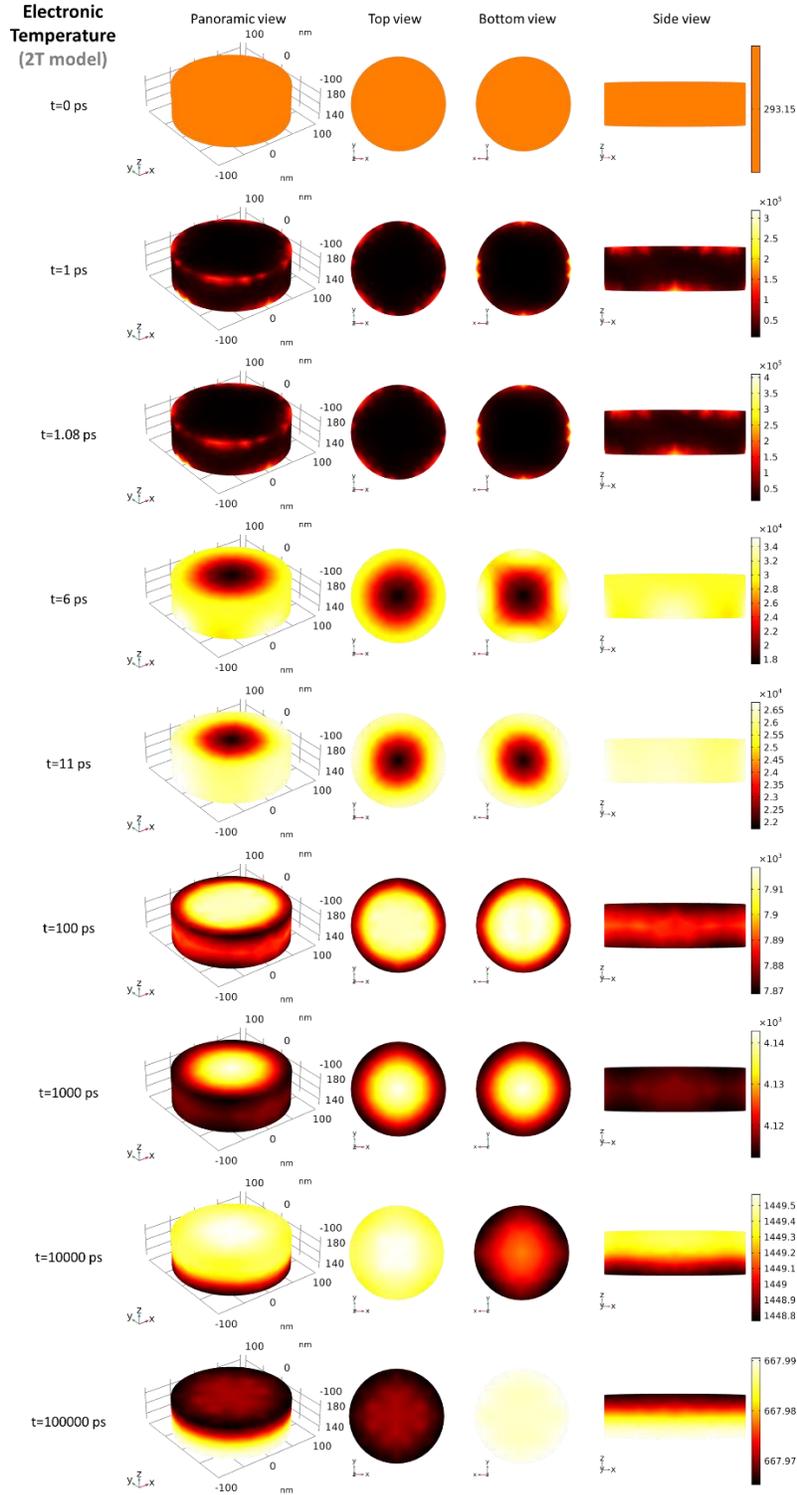

**Figure S28:** Electronic temperature heat maps for the disk antenna in the metastructure configuration, for the solutions of the 2T model with LCP pump. The metastructure was simulated with $I_0 = 10^{12}$ W/cm$^2$, with an LCP beam $\vec{E}//-k_z$. All bar units are in K.

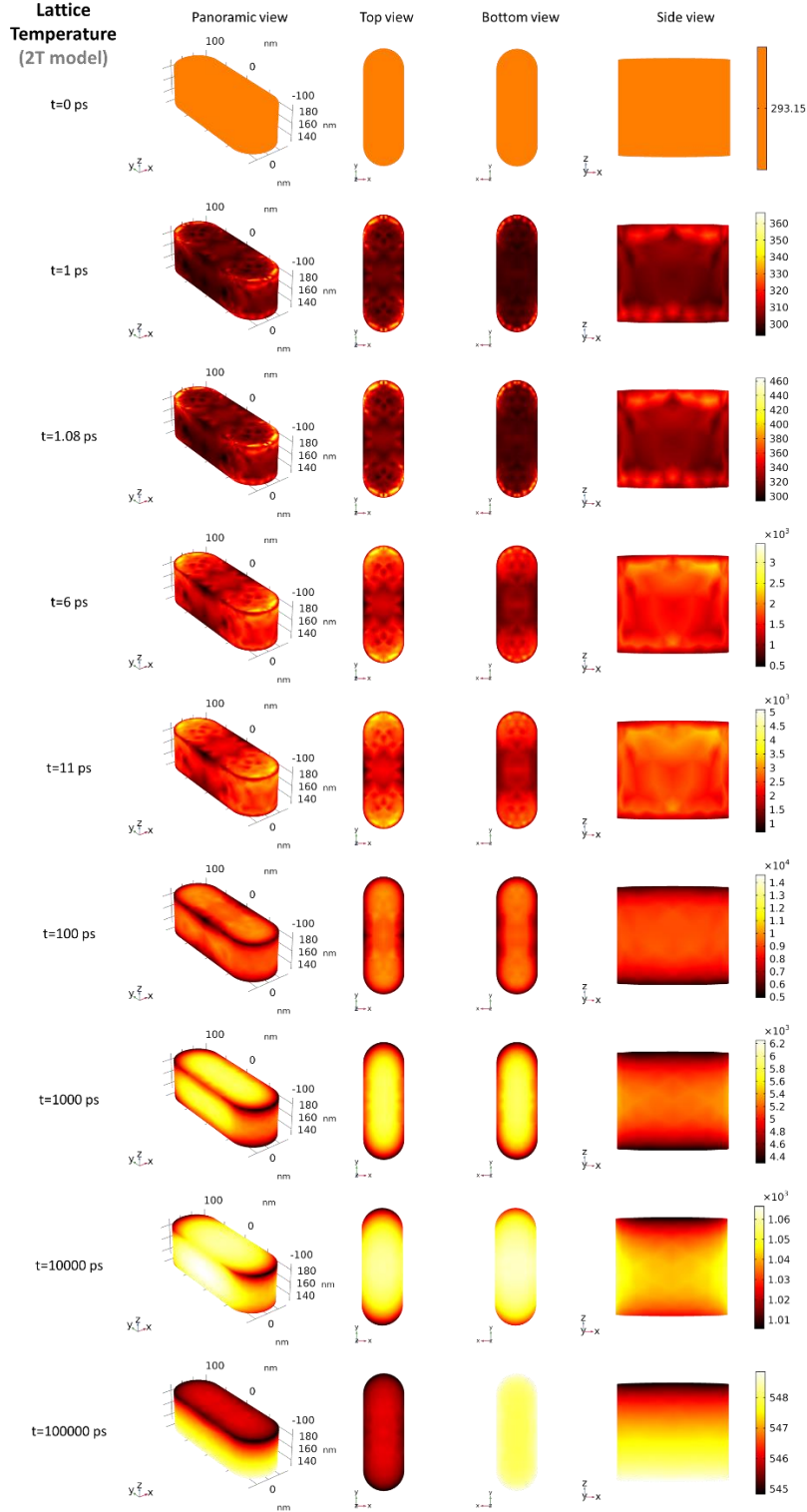

**Figure S29:** Lattice temperature heat maps for the rod antenna in the metastructure configuration, for the solutions of the 2T model with LCP pump. The metastructure was simulated with $I_0 = 10^{12}$ W/cm$^2$, with an LCP beam $\vec{E}\,//-k_z$. All bar units are in K.

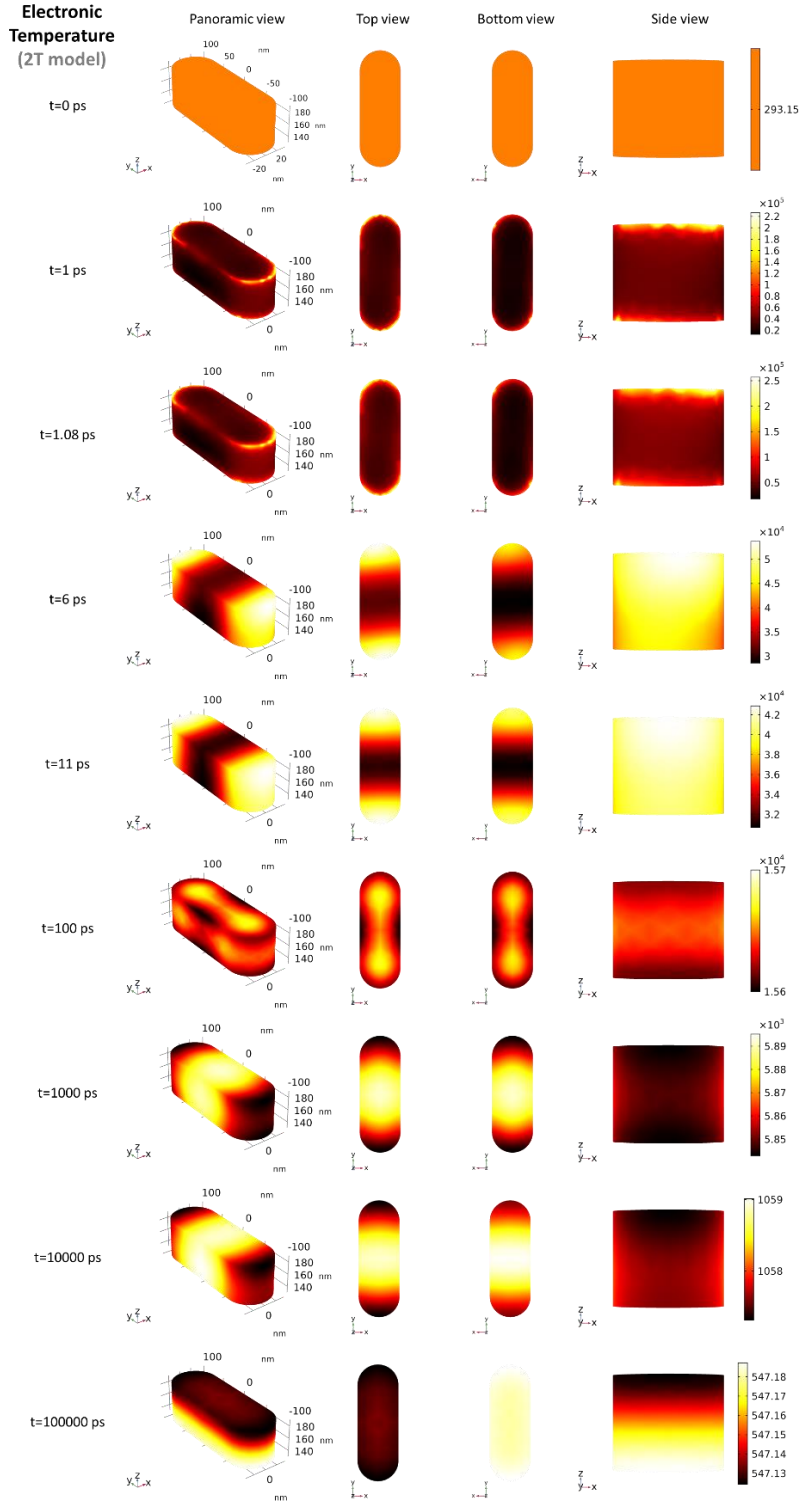

**Figure S30:** Electronic temperature heat maps for the rod antenna in the metastructure configuration, for the solutions of the 2T model with LCP pump. The metastructure was simulated with $I_0 = 10^{12}$ W/cm$^2$, with an LCP beam $\vec{E}\,//-k_z$. All bar units are in K.

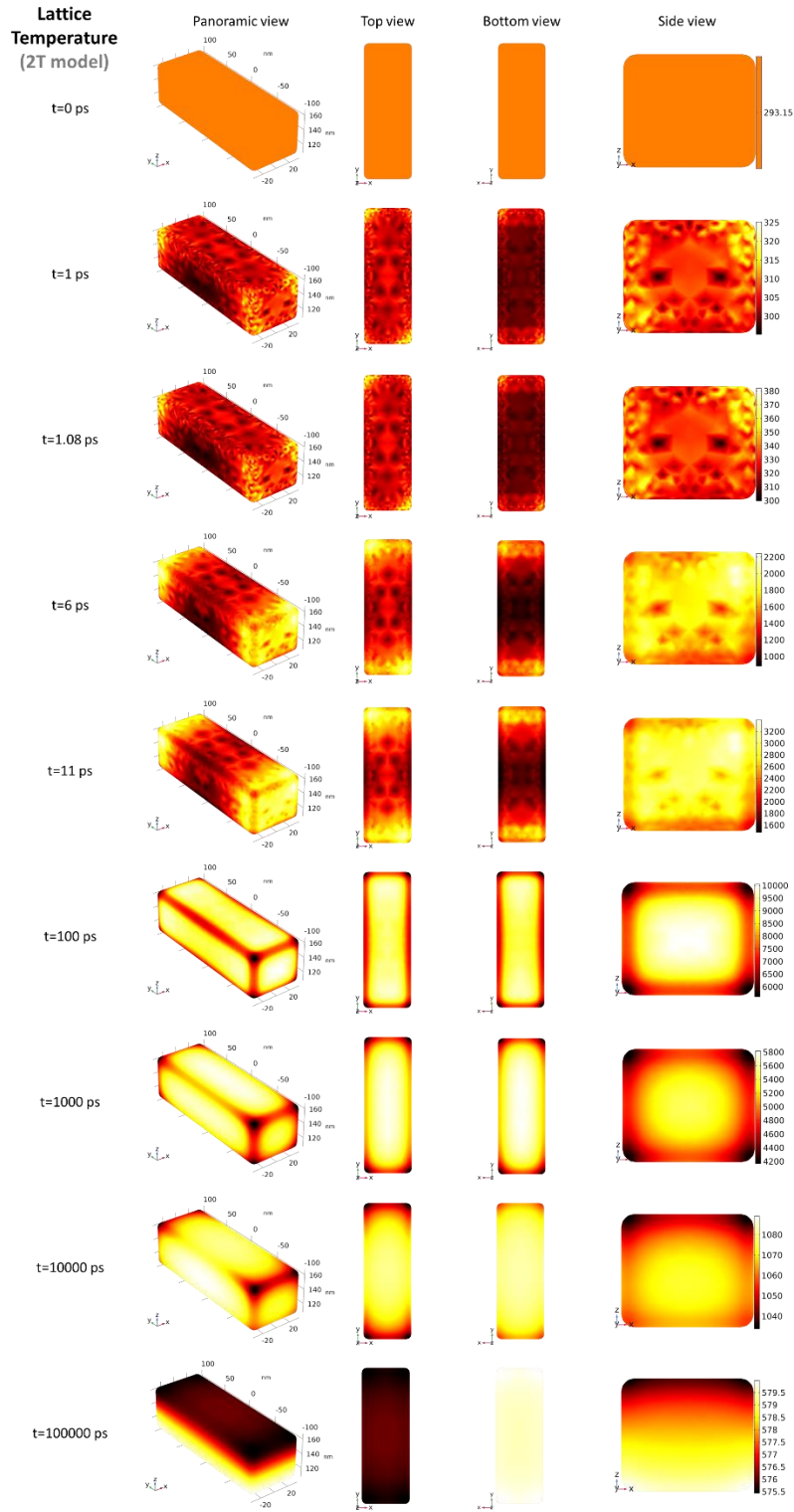

**Figure S31:** Lattice temperature heat maps for the rectangular prism antenna in the metastructure configuration, for the solutions of the 2T model with LCP pump. The metastructure was simulated with $I_0 = 10^{12}$ W/cm$^2$, with an LCP beam $\vec{E}\,//-k_z$. All bar units are in K.

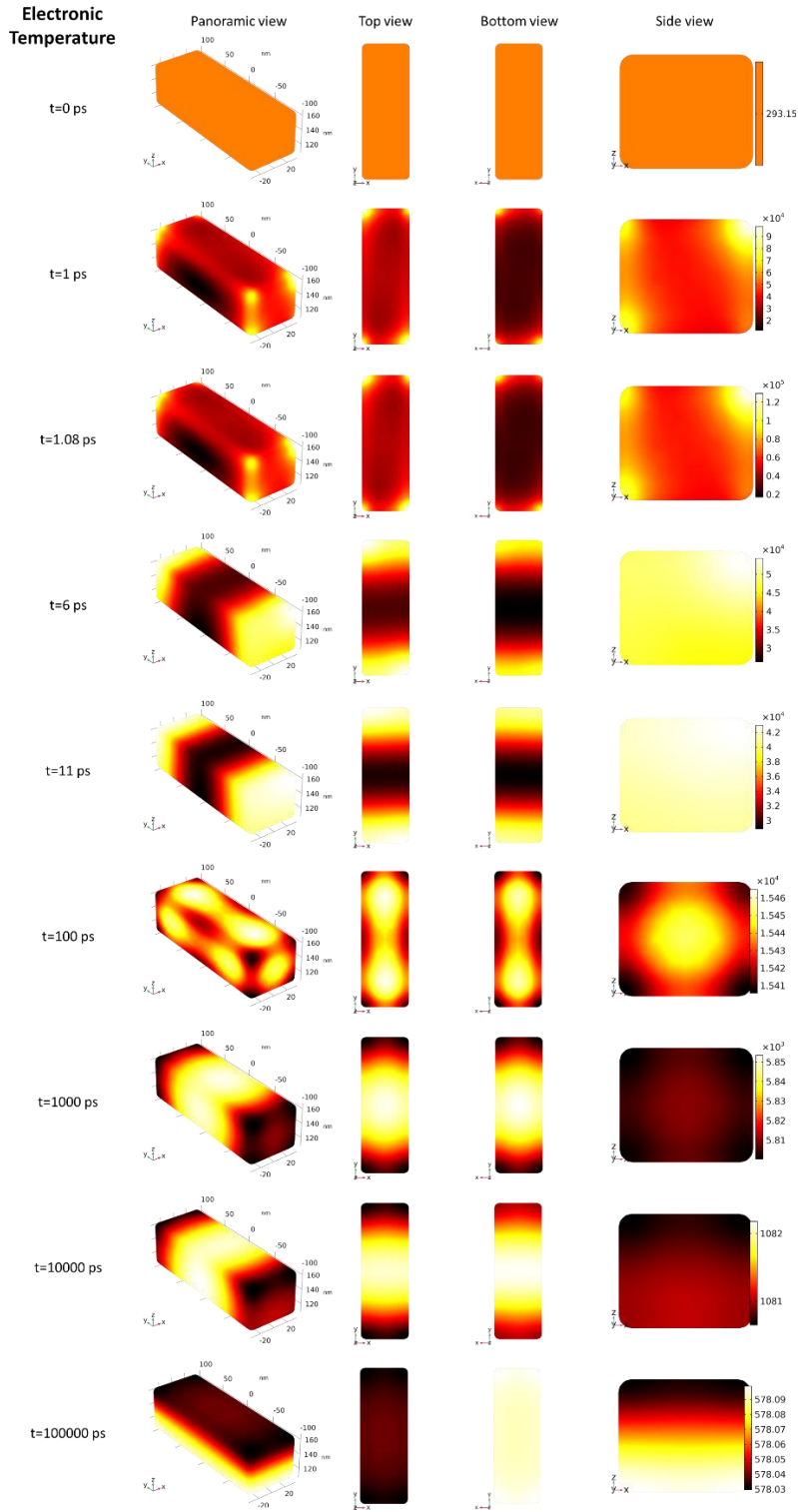

**Figure S32:** Electronic temperature heat maps for the rectangular prism antenna in the metastructure configuration, for the solutions of the 2T model with LCP pump. The metastructure was simulated with $I_0 = 10^{12}$ W/cm$^2$, with an LCP beam $\vec{E} // -k_z$. All bar units are in K.

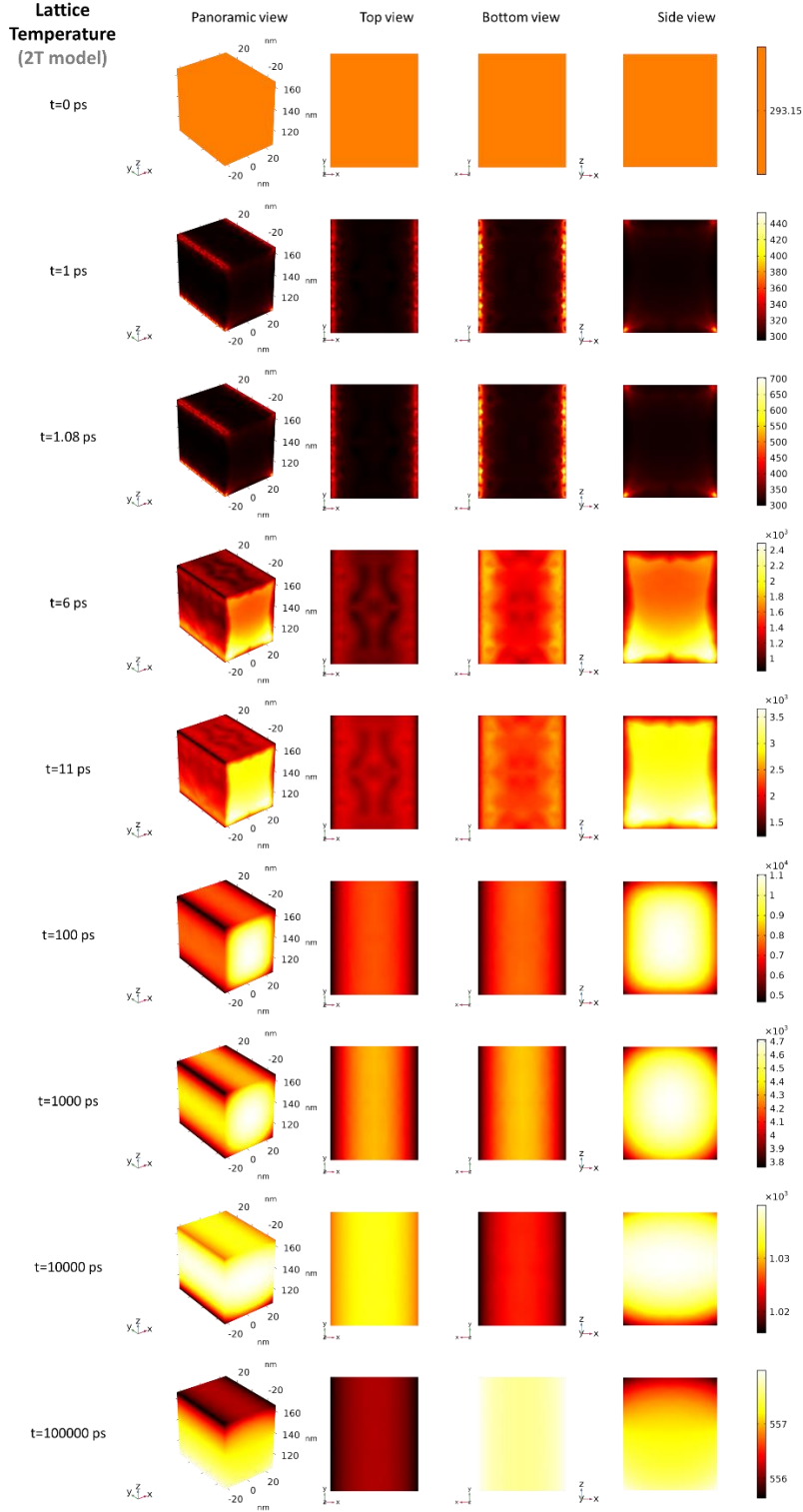

**Figure S33:** Lattice temperature heat maps for the achiral wire antenna in the metastructure configuration, for the solutions of the 2T model with LCP pump. The metastructure was simulated with $I_0 = 10^{12}$ W/cm$^2$, with an LCP beam $\vec{E}\,//-k_z$. All bar units are in K.

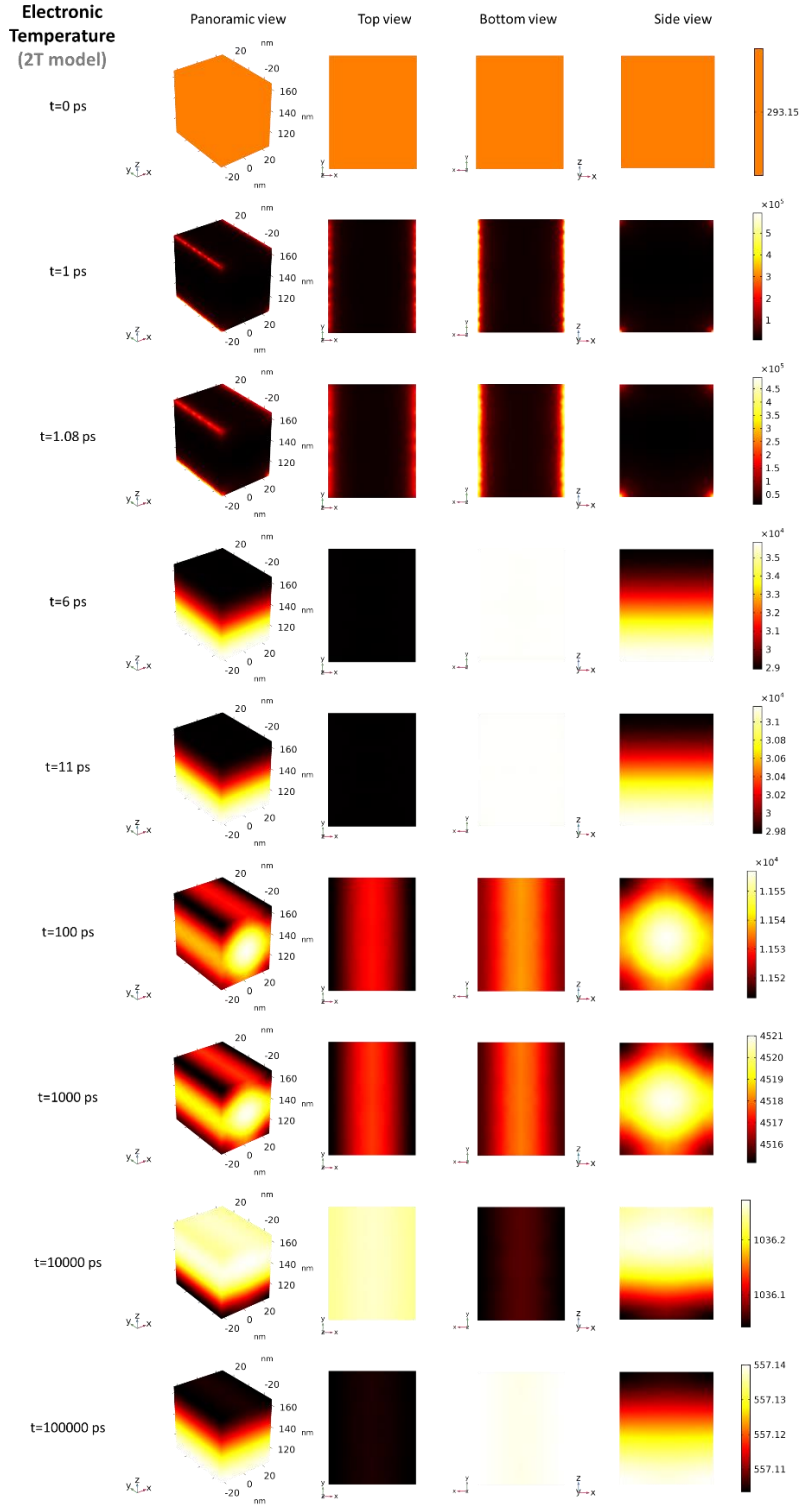

**Figure S34:** Electronic temperature heat maps for the achiral wire antenna in the metastructure configuration, for the solutions of the 2T model with LCP pump. The metastructure was simulated with $I_0 = 10^{12}$ W/cm$^2$, with an LCP beam $\vec{E}\,//-k_z$. All bar units are in K.

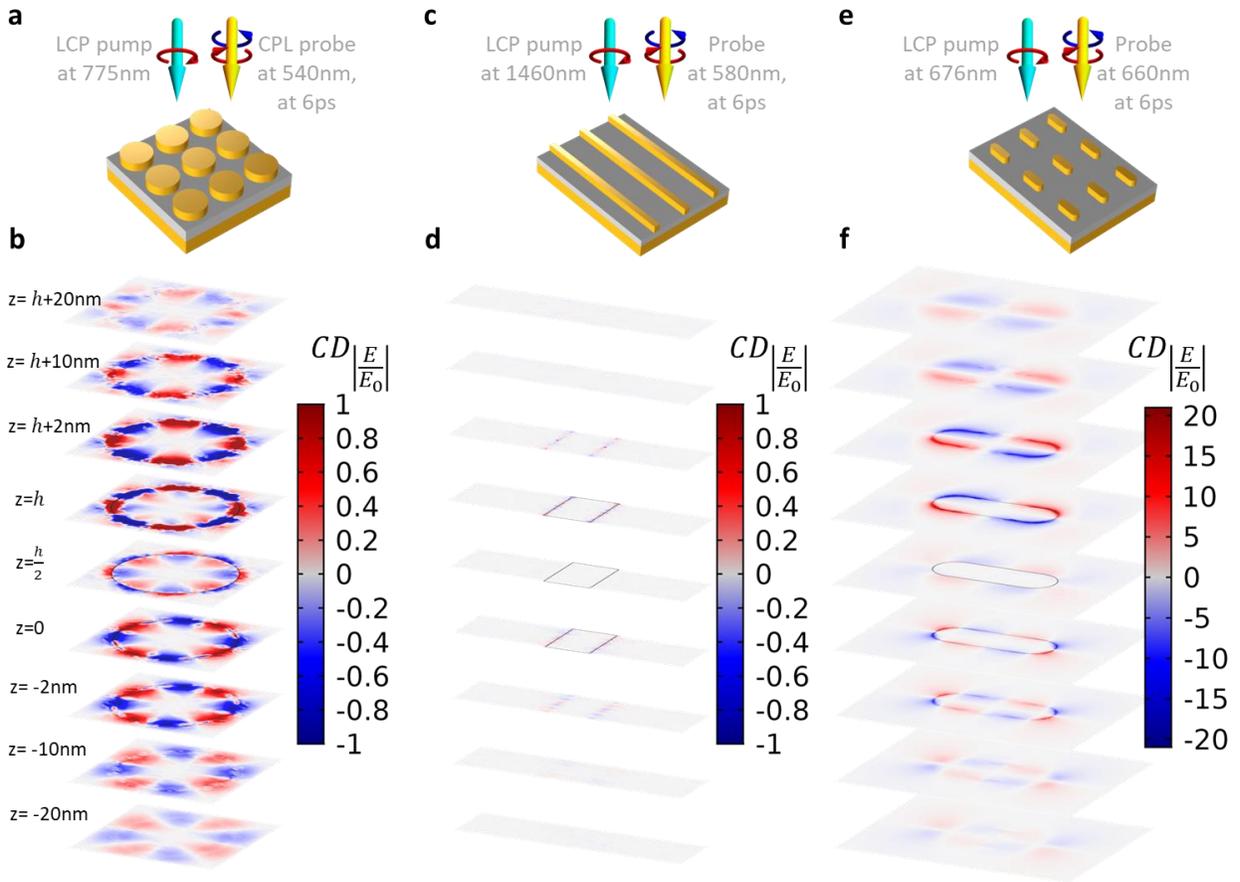

**Figure S35:** CD of the electric field distributions for the disk (a,b), for the achiral wire (c,d), and for the rod (e,f), all at the times t=5ps, this is when each structure shows a maximum absorption CD in Figs. 5h, 5h and and 5e respectively, within the 2T-model. For the rest of the structures see Fig. 6. Each plane is a two-dimensional cut as indicated at the panel b, where $h$ is the height of each antenna, as indicated in Figs. S6 and S7, such that $z = 0$ corresponds to the bottom of each Au antenna.

**Discussion on the chiral time-dynamic in various metastructures**

In general, for the chiral wire we see a larger thermal response for LCP pump. Whereas the 1T model shows a similar behavior as in the spherical Au NPs (Figs. S3 and S4), meaning a sudden increase of the phonon temperature, here the 2T model shows a large difference. The $T_e$ rises at its maximum ~0.36 ps after irradiation, and the $T_l$ at ~36 ps after irradiation, with an electron-phonon relaxation times of 9.64 ps. Also, $T_e$ does not follow the typical exponential decay, but a more complex behavior. For the L-shape, we see a larger thermal response for RCP pump. Again, the 2T model is the most notably different: $T_e$ rises at its maximum ~1 ps after irradiation, and the $T_l$ at ~100 ps after irradiation, with an electron-phonon relaxation times of 7.52 ps. Te does not follow the typical exponential decay. In agreement with the CW responses, here the chiral wire shows much larger temperatures for LCP pump than RCP pump (see Fig. 2d), whereas the L-shaped shows a slightly larger RCP pump temperature than LCP (see Fig. 2a). The spatial distributions of the temperatures $T_l$ and $T_e$, respectively, are shown in Figs. S20 and S21 for the chiral wire, and S22 and S23 for the L-shaped. For the chiral wire, we can see that the electrons are largely excited at the wire's juncture (Fig. S21) and then dissipating their energy throughout the rest of the structure simultaneously interacting with the phonons (Fig. S20) which are excited at later times as compared to the electrons. Similarly, the L-shaped arms are heated up for electrons at the middle of both of its arms (Fig. S22), allowing these phonons to be excited a later times (Fig. S23), and then distributing the heat throughout the arms.

Lastly, in Fig. S35 we show the spatial distribution maps of the photo-induced CD with the pump-probe setup, for the rest of achiral metastructures not shown in Fig. 6. These maps are calculated for the new shifted resonances after pumping and for the times when the $CD_{abs}(t)$ curves are maximum within the 2T model, shown in Figs. 5e and 5h. We can see that the maps show positive (red) and negative (blue) regions, which can be excited with LCP and RCP, respectively, as demanded. A distinct difference is that whereas for the chiral structures, these regions are always accessible at every time (their intensity will be different though), for the achiral structures excited opposite corner chiral plasmons will be accessible only at very specific times. For the CD maps of the achiral wire, no imprinted chirality is notable at any part of the structure, as expected. The rectangular prism (Fig. 6f) and the rod (Fig. S35f), show that the structure exhibits more CD intensity at the top of the antenna than at the bottom, and also that whereas the top excitation is a quadrupole the bottom excitation is an octupole. This difference in intensity and pole distribution is what yields a finite probe-CD, phenomena we termed as chirality imprinting, which only lasts a few ps after irradiation.